\documentclass{PoS}

\usepackage{graphicx}
\usepackage{amsmath}
\usepackage{amsfonts}
\usepackage{subfigure}
\usepackage{wrapfig}
\usepackage{epsfig}
\usepackage{afterpage,float}

\newcommand{\be}{\begin{equation}}
\newcommand{\ee}{\end{equation}}
\newcommand{\bea}{\begin{eqnarray}}
\newcommand{\eea}{\end{eqnarray}}

\def\half{{\textstyle{1\over2}}}

\def\bar{\overline}

\def\tilde{\widetilde}

\def\half{{\scriptstyle \raise.15ex\hbox{${1\over2}$}}}

\newcommand{\beq}{\begin{equation}}
\newcommand{\eeq}{\end{equation}}

\newcommand{\real}{\relax{\rm I\kern-.18em R}}

\title{Chiral symmetry breaking in fundamental and sextet fermion representations of SU(3) color
\thanks{Based on talks at the 
conference by J.~Kuti and Kieran Holland. }}

\ShortTitle{Chiral symmetry breaking in fundamental and sextet fermion representations of SU(3) color}

\author{Zolt\'an Fodor\\
        Department of Physics, University of Wuppertal\\
        Gau$\beta$strasse 20, D-42119, Germany\\
        J\"ulich Supercomputing Center, Forschungszentrum\\
        J\"ulich, D-52425 J\"ulich, Germany\\
        Email: \email{fodor@bodri.elte.hu}}

\author{Kieran Holland\\
        Department of Physics, University of the Pacific\\
        3601 Pacific Ave, Stockton CA 95211, USA\\
        Email: \email{kholland@pacific.edu}}

\author{Julius Kuti \\
        Department of Physics 0319, University of California, San Diego\\
        9500 Gilman Drive, La Jolla, CA 92093, USA\\
        E-mail: \email{jkuti@ucsd.edu}}

\author{D\'aniel N\'ogr\'adi\\
        Institute for Theoretical Physics, E\"otv\"os University\\
        H-1117 Budapest, Hungary\\
        Email: \email{nogradi@bodri.elte.hu}}

\author{Chris Schroeder\\
        Department of Physics, University of Wuppertal\\
        Gau$\beta$strasse 20, D-42119, Germany\\       
        E-mail: \email{chris.schroeder@gmail.com}}

\abstract{We report new results for lattice gauge theories with twelve fermion flavors  
in the fundamental  representation and two fermion flavors 
in the two-index symmetric (sextet) representation
of the SU(3) color gauge group. Both models are important in searching for a viable composite 
Higgs mechanism in the Beyond the Standard Model (BSM) paradigm. 
We subject both models to opposite hypotheses inside and outside of
the conformal window. In the first hypothesis we test  chiral symmetry breaking ($\chi{\rm SB}$) with
its Goldstone spectrum, $F_\pi$,  the $\chi{\rm SB}$ condensate, 
and several composite hadron states as the fermion mass is varied
in a limited range with our best effort to control finite volume effects.
Supporting results for $\chi{\rm SB}$ from the running coupling based on the force between static sources 
is also presented.
In the second test for the alternate hypothesis we probe conformal behavior driven by a single anomalous mass dimension
under the assumption of unbroken chiral symmetry. Our  results show very low level of confidence
in the conformal scenario. 
\vskip 0.05 in}

\FullConference{The XXVIII International Symposium on Lattice Field Theory, Lattice2010\\
		June 14-19, 2010\\
		Villasimius, Italy}

\begin{document}

\section{Introduction}

We report $\chi{\rm SB}$ studies of two important gauge theories which attracted a great deal 
of attention in the lattice community and off-lattice as well.
To establish the chiral properties of a gauge theory close to the conformal window is notoriously difficult.
If the chiral symmetry is broken, the fundamental parameter $F$ of the chiral Lagrangian has to be small in 
lattice units $a$  to control cut-off effects. Since the chiral expansion has terms with powers 
of  $N_f M^2_\pi/16\pi^2 F^2$, reaching the chiral regime with large number of fermion flavors is
particularly difficult. 
The range of $aM_\pi$ values where 
leading chiral logs can be identified unambigously will require simulations in very large volumes 
which is not in the scope of this study with twelve fermion flavors in the fundamental representation of the
$SU(3)$ color gauge group.
The sextet representation with two fermion flavors  is considerably
closer to the range of chiral perturbation theory  in our simulations. 
Consistency requirements of chiral logs which fit the sextet results will require to get closer
to the loop expansion of the continuum chiral Lagrangian, or the comprehensive application of staggered $SU(2)$ chiral
perturbation theory on coarser lattices.
We will make a case in this report that 
qualitatively different expectations inside and outside of  the conformal window allow
 tests of the two mutually exclusive hypotheses without reaching down to the chiral logs at very small pion masses. 

Below the conformal window
chiral symmetry is broken at zero fermion mass with a gap in the composite hadron spectrum 
except the associated massless Goldstone multiplet. 
The analytic form of the chiral Lagrangian as a function of the
fermion mass can be used to detect chiral log corrections, or to differentiate from conformal exponents in the transitional
region before the chiral logs are reached at low enough Goldstone pion masses.  
Approximations to gauge theories with  $\chi{\rm SB}$, 
like their effective Nambu-Jona-Lasinio 
description in the large N limit, are consistent with this analysis. 
In sharp contrast, the spectrum inside the conformal window is gapless in all channels 
in the chiral limit and the scale dependence of physical quantities is governed by the single 
critical exponent $\gamma$ which controls the fermion mass dependence of 
composite operators and their correlators. 

The two competing hypotheses are  tested for
both gauge theories studied here in search for their chiral properties. 
%It is important to emphasize a rather powerful argument here.
There is a fundamental difference between the two hypotheses as implied by their respective
spectra.  $\chi{\rm SB}$ creates 
a fundamental scale $F$ in the theory separated from the composite hadron scale with 
its  residual baryon gap in the chiral limit.
The pion mass can be varied from the $\chi{\rm SB}$ scale $F$ to the hadron scale with a transition
from the chiral log regime to a regime without chiral analysis. The conformal phase has no
intrinsic scale.
With  $\chi{\rm SB}$ this is expected to lead to fermion mass dependence of the spectrum 
in the chiral log regime, or above it, quite different from
the conformal behavior which is very tightly constrained near the chiral limit 
of the spectrum with a single critical exponent
$\gamma$ in the absence of any intrinsic scale.
In a regime where lattice cutoff effects are negligible, this difference 
should be sufficient for  tests whether the chiral
loop expansion is reached, or not, on the low $F$ scale.

In Section 2 we present new results for the gauge  model with twelve fermions in the fundamental representation.
A new kind of gauge dynamics 
is expected to appear at intermediate distances with 
walking gauge coupling, or a conformal fixed point. This remained controversial with recent efforts 
from five lattice groups~\cite{Appelquist:2007hu, Fodor:2009wk, Deuzeman:2009mh, Jin:2009mc,Hasenfratz:2010fi}.
We made considerable progress to resolve the controversies including tests of the chiral
condensate and the spectrum which favor chiral symmetry breaking with unusual chiral dynamics. 
Applying the $\chi{\rm SB}$ 
 hypothesis to the Goldstone pion,  $F_\pi$, the chiral condensate, and the stable nucleon state collectively leads to
a result of ${\rm \chi^2/dof=1.22}$ representing high level of confidence. 
With the conformal hypothesis  we find ${\rm \chi^2/dof=8.79}$ representing very low level of confidence.
Applying global analysis to all states we measured, the contrasting behavior is somewhat less dramatic but remains significant.
New results on the running coupling from the static force  and our 
simulation of a rapid finite temperature transition in Polyakov loop distributions,  reported elsewhere
and  expected  in association with 
$\chi{\rm SB}$ and its restoration, provide further support for our findings.

%Although our new results should not be viewed as definitive, they are quite powerful.
%In all fits we were on a fine-grained lattice in the pion mass range $aM_{\pi}=0.16-0.39$ 
%and rho mass range $M_{\rho}=0.2-0.47$. In contrast, a previous study~\cite{Deuzeman:2009mh}
%which reported conformal behavior was in the $aM_{\pi}=0.35-0.67$ range and rho mass range $M_{\rho}=0.39-0.77$.

In Section 3 we report new results for the two-index symmetric (sextet) color representation with two fermion flavors. 
The model holds  promise for the composite Higgs mechanism in its simplest implementation without unwanted 
extra Goldstone bosons, or new electroweak multiplets. If  $\chi{\rm SB}$ is found, the model is like the textbook 
introduction to Technicolor from QCD except that being close to the conformal window it 
becomes a viable BSM alternative without Electroweak precision problems. 
In a similar analysis introduced above, the sextet model was also subjected to the two mutually exclusive hypotheses. 
Applying the $\chi{\rm SB}$ 
 hypothesis to the Goldstone pion,  $F_\pi$, and the chiral condensate,
a result of ${\rm \chi^2/dof=1.24}$ was found representing again high level of confidence. 
With the conformal hypothesis  we find ${\rm \chi^2/dof=6.96}$ representing very low level of confidence.

We have used the tree-level Symanzik-improved gauge action for all simulations in this paper.
The conventional $\beta=6/g^2$ lattice gauge coupling is defined as the overall
factor in front of the well-known terms of the Symanzik lattice action.  Its value is $\beta=2.2$ for all simulations
of the $N_f=12$ model and $\beta=3.2$ in the sextet model. The link
variables in the staggered fermion matrix were exponentially smeared with  two
stout steps~\cite{Morningstar:2003gk} and the precise definition of the action is given in~\cite{Aoki:2005vt}.  
The RHMC and HMC algorithms were deployed in all runs.
Our error analysis of effective mass plots which combines systematic 
and statistical effects follows the frequentist histogram approach 
of the Budapest-Marseille-Wuppertal collaboration~\cite{ Durr:2010aw}
in all simulations. 
%The 68 percent width of the histogram of all effective mass fits, weighted with their Q-values for
%fits starting at $N_t/4$ on the plateaux, was taken as the error in each channel. 
The topological charge was monitored
in the simulations with frequent changes observed over a considerable range.

\section{Twelve fermions in the fundamental SU(3) color representation}

The chiral Lagrangian for the Goldstone spectrum separated from the massive composite 
scale of hadrons exhibits, order by order, the well-known analytic form of  powers in the fermion mass $m$  
with non-analytic chiral  log corrections 
generated from pion loops close enough to the chiral limit. 
The exact  functions $F_\pi(m)$ and $M_\pi(m)$ will be approximated by an analytic form in powers of
$m$ which is expected to hold over a limited $m$ range when the Goldstone pion is in transition from
 the chiral log regime  closer to the composite hadron scale. Although this procedure has some
 inherent uncertainty before the chiral logs are reached in simulations, its sharp contrast with the 
 non-analytic fermion mass dependence of the conformal hypothesis, governed by the single exponent $\gamma$, is
 sufficient to differentiate the two hypotheses.
 
First, we will illustrate the fitting procedure with results on the Goldstone spectrum, $F_\pi$, and the chiral 
condensate. This will be extended  
to the nucleon and some other composite hadron channels to probe parity degeneracy in the chiral limit.
\subsection{Goldstone spectrum and $\bf F_\pi$ from  chiral symmetry breaking}
\begin{figure}[!htpb]
\begin{center}
\begin{tabular}{ccc}
\includegraphics[height=4cm]{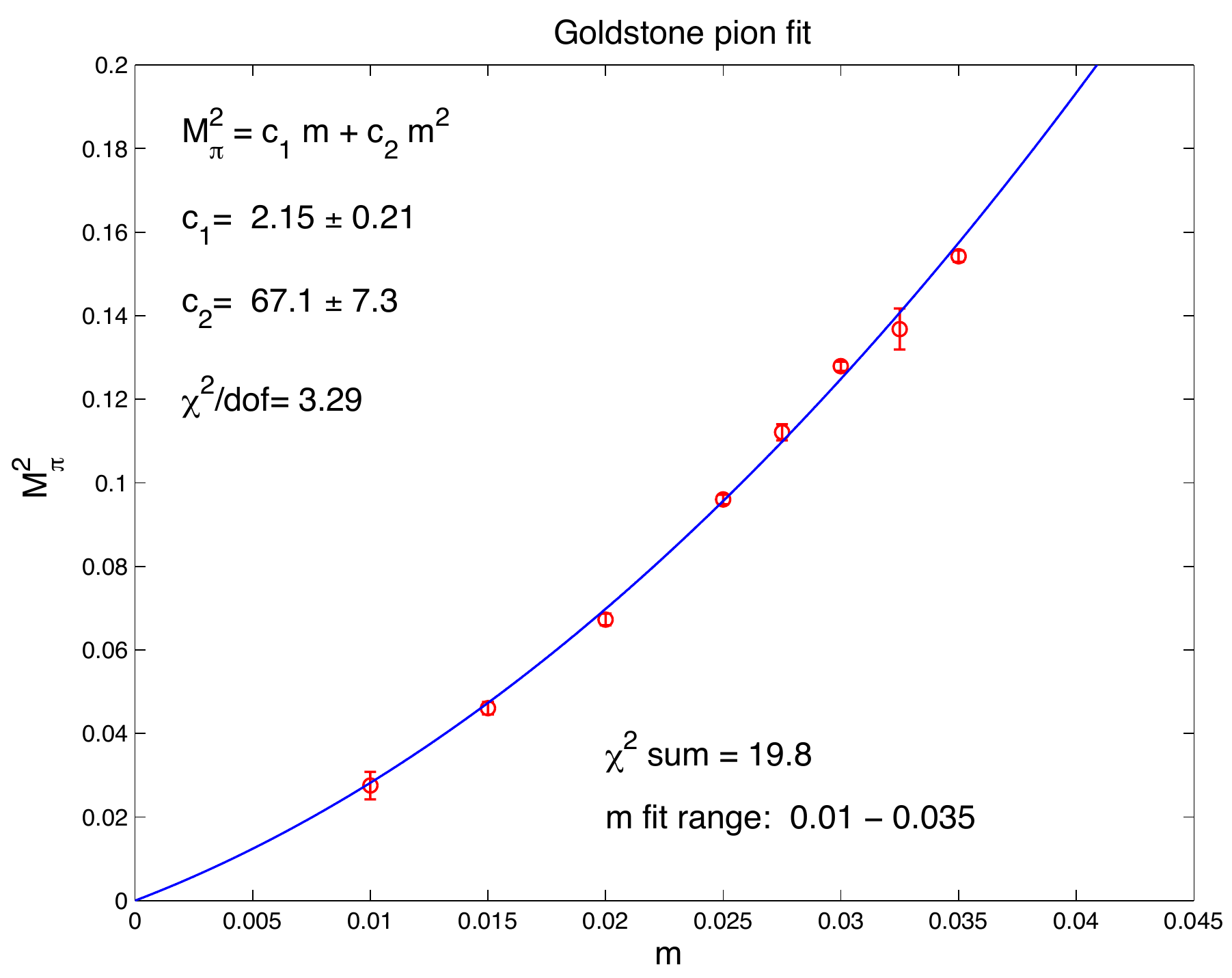}&
\includegraphics[height=4cm]{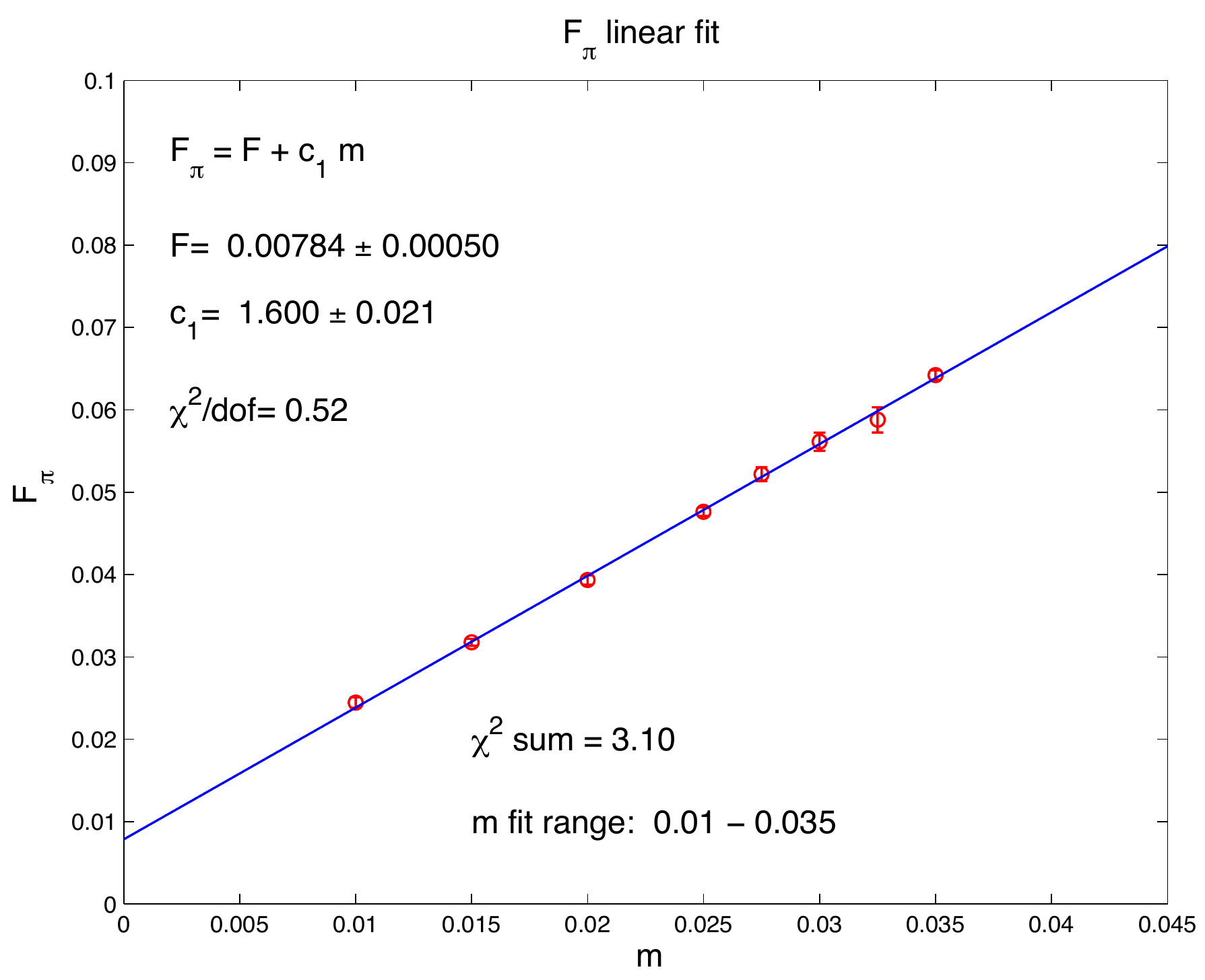}\\
\includegraphics[height=4cm]{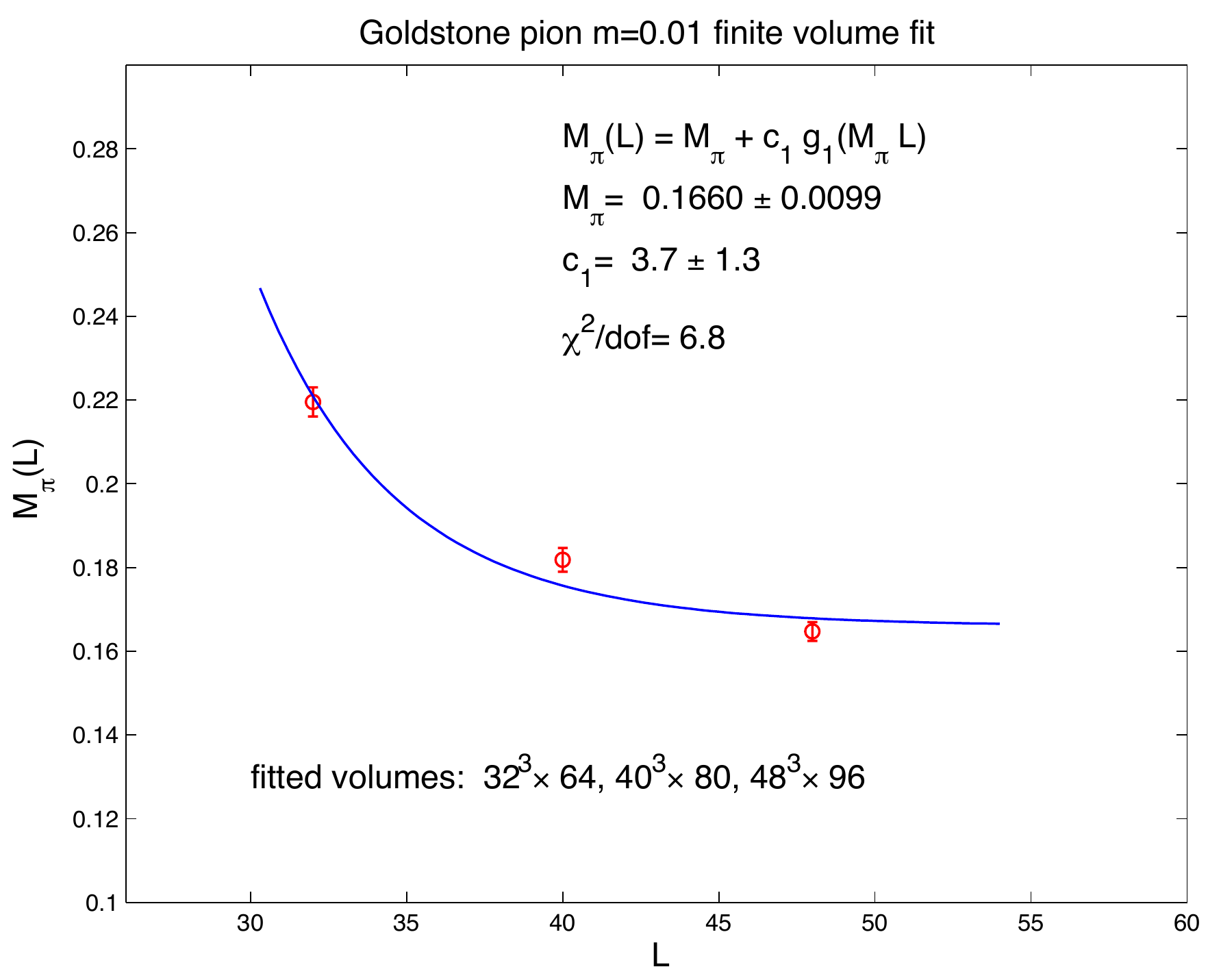}&
\includegraphics[height=4cm]{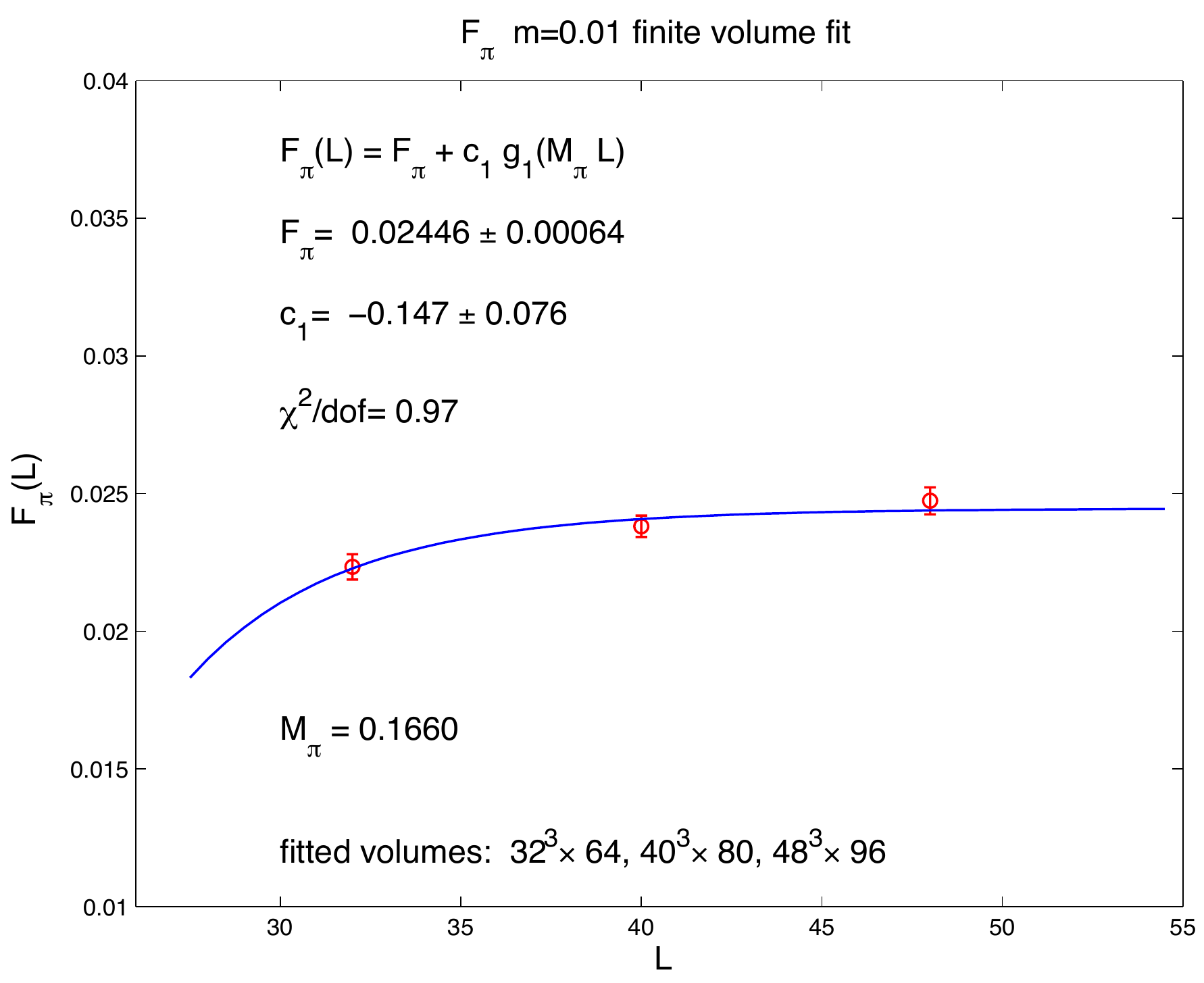}
\end{tabular}
\caption{{\footnotesize The Goldstone pion and $ F_\pi$ from  
chiral symmetry breaking are shown with the fitting procedure described in the text.
Representative finite volume fits are also shown. The infinite volume limit of $M_\pi$ was used in fits to $F_\pi$ and other composite
hadron states, like the nucleon.}}
\end{center}
\label{fig:Fpi-Pion}
\vskip -0.2in
\end{figure}

Figure 1 shows the Goldstone pion and $F_\pi$ as a function of the fermion mass $m$ in the range where we can
reach the infinite volume limit with  confidence.
The power functions of the fitting procedure in $m$ contain the analytic contributions of the fourth order 
chiral Lagrangian to $M_\pi$ and $F_\pi$.  Although we could fit the pion spectrum with the 
logarithmic term included, its significance remains unclear.
The rapid variation of $F_\pi$ with $m$ clearly shows that we would need 
a dense set of data in the $m=0.003-0.01$ range to reach chiral logs at this gauge coupling. 
This requires lattice volumes
well beyond the largest size $48^3\times 96$ which we could deploy in our simulations. 

Efforts were made for extrapolations to the infinite volume limit. 
Finite volume scaling is very different under the hypotheses of the two different scenarios. 
At the lowest three $m$ values, for  finite volume corrections to  $M_\pi$ and $F_\pi$, and for 
all other states, we used the form
\begin{eqnarray}
 M_\pi(L_s,\eta)& = &M_\pi  \biggl [1+\frac{1}{2N_f}\frac{M^2}{16\pi^2F^2}\cdot\tilde g_1(\lambda,\eta) \biggr ] ,
 \nonumber\\
     F_\pi (L_s,\eta) &= &F_\pi \biggl [1-\frac{N_f}{2}\frac{M^2}{16\pi^2F^2} \cdot\tilde g_1(\lambda,\eta) \biggr ] ,      
\label{eq:MpiL}
\end{eqnarray}
where $\tilde g_1(\lambda,\eta)$ describes the finite volume corrections with
$\lambda=M_\pi\cdot L_s$ and aspect ratio $\eta=L_t/L_s$ from the lightest pion wrapping around the lattice
and coupled to the measured state.  
The form of $\tilde
g_1(\lambda,\eta)$ is a complicated infinite sum which contains Bessel
functions and requires numerical evaluation. Since we are not in the chiral log regime, the pre-factor of
the $\tilde g_1(\lambda,\eta)$ function was replaced by a fitted coefficient. The leading term of  the function
$\tilde g_1(\lambda,\eta)$ is a special exponential Bessel function $K_1(\lambda)$ which dominates in the simulation range.
The fitting procedure could be viewed as the approximate  leading treatment of  the pion which wraps around the finite volume,
whether in chiral perturbation theory, or in Luscher's non-perturbative finite volume analysis.
The $M_\pi L_s > 4$ lore for volume independence is clearly not applicable in the
model. We need $M_\pi L_s > 8$ to reach volume independence.
The infinite volume limits of $M_\pi$ and $F_\pi$  for each $m$ were determined self-consistently from the fitting
procedure using Eqs. (2.1) based on 
a set of $L_s$ values with  representative fit results shown in Figure 1. In the higher $m$ range finite volume effects
were hard to detect and even for the lowest $m$ values sometimes volume dependence was not detectable for the largest
lattice sizes. 

Non-Goldstone pion spectra, quite different from those found in QCD, are shown in Figure 2 using standard notation.
They are not used in our global analysis.
The three states we designate as i5Pion,
ijPion and scPion do not show any noticeable taste breaking or residual mass in the $m\rightarrow 0$ chiral limit. 
The scPion is degenerate with the i5Pion and both are somewhat split from the true Goldstone pion. 
The ijPion state is  further split as expected but the overall taste breaking is very small across four pion states. 
This is a fairly strong indication that the coupling 
constant $\beta=2.2$ where all runs are performed is close to the continuum limit.
A very small residual mass at $m=0$ is not excluded for some non-Goldstone states depending on the
details of the fitting procedure. 

\begin{figure}[htpb]
\begin{center}
\begin{tabular}{cc}
\includegraphics[height=4cm]{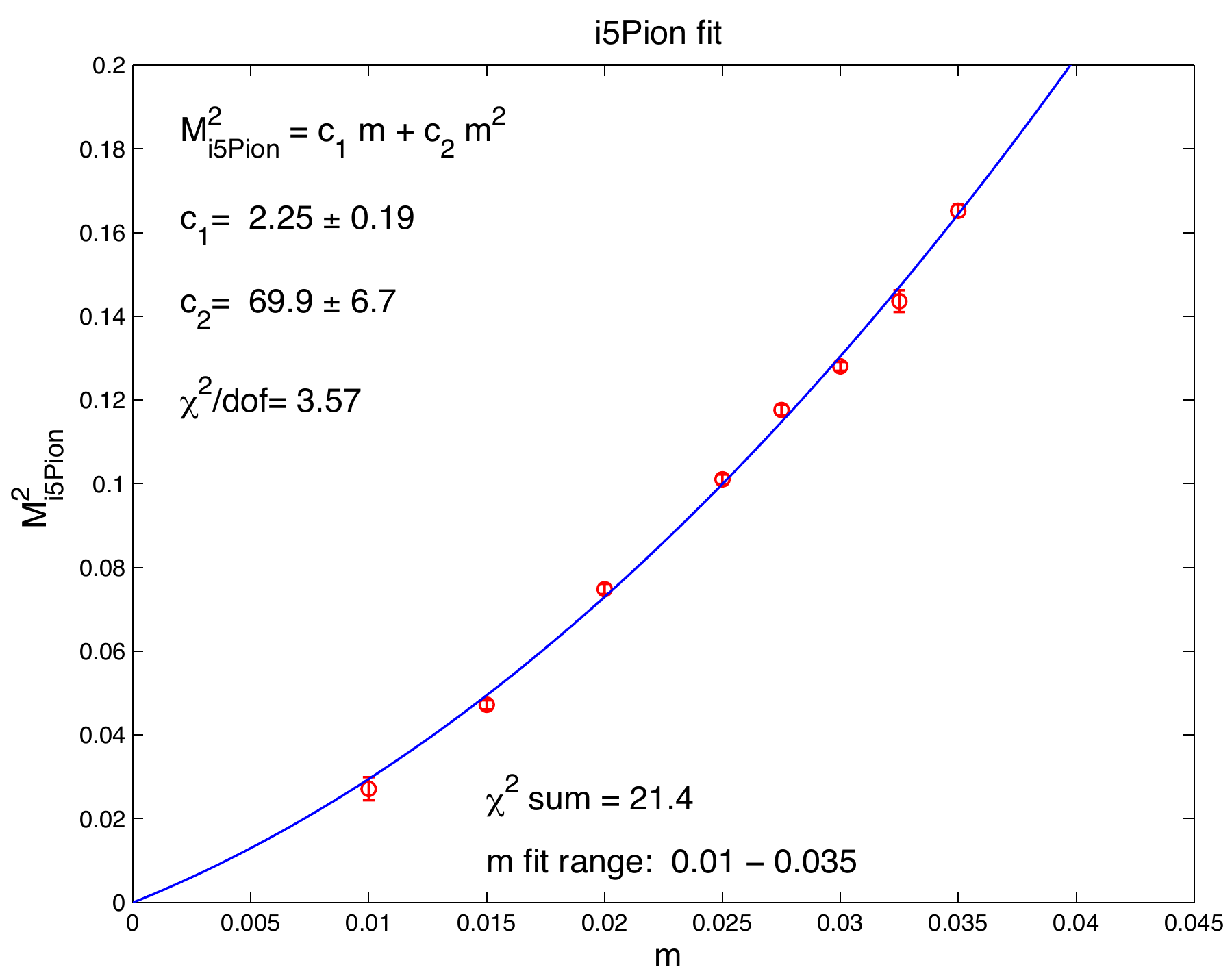}&
\includegraphics[height=4cm]{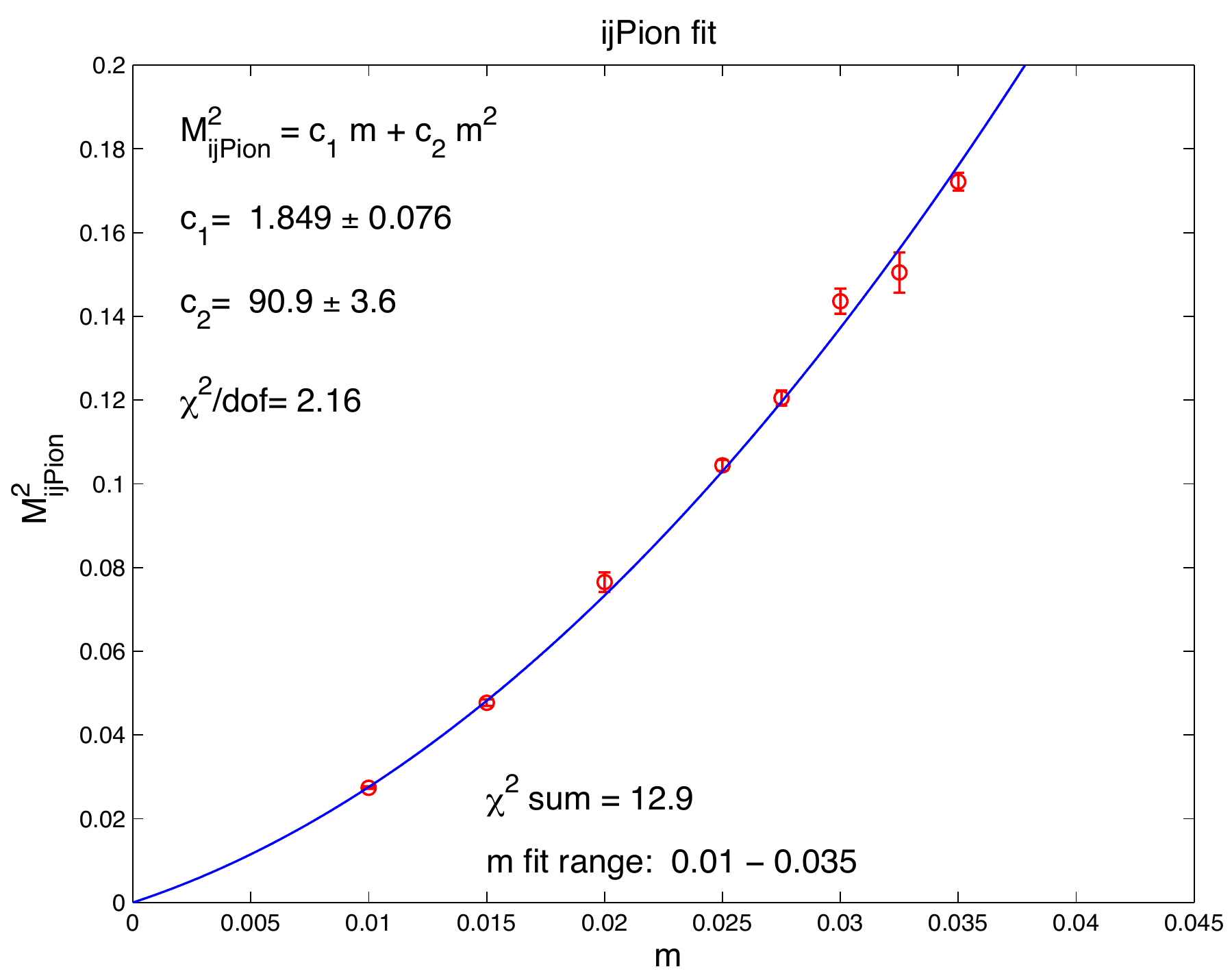}\\
\includegraphics[height=4cm]{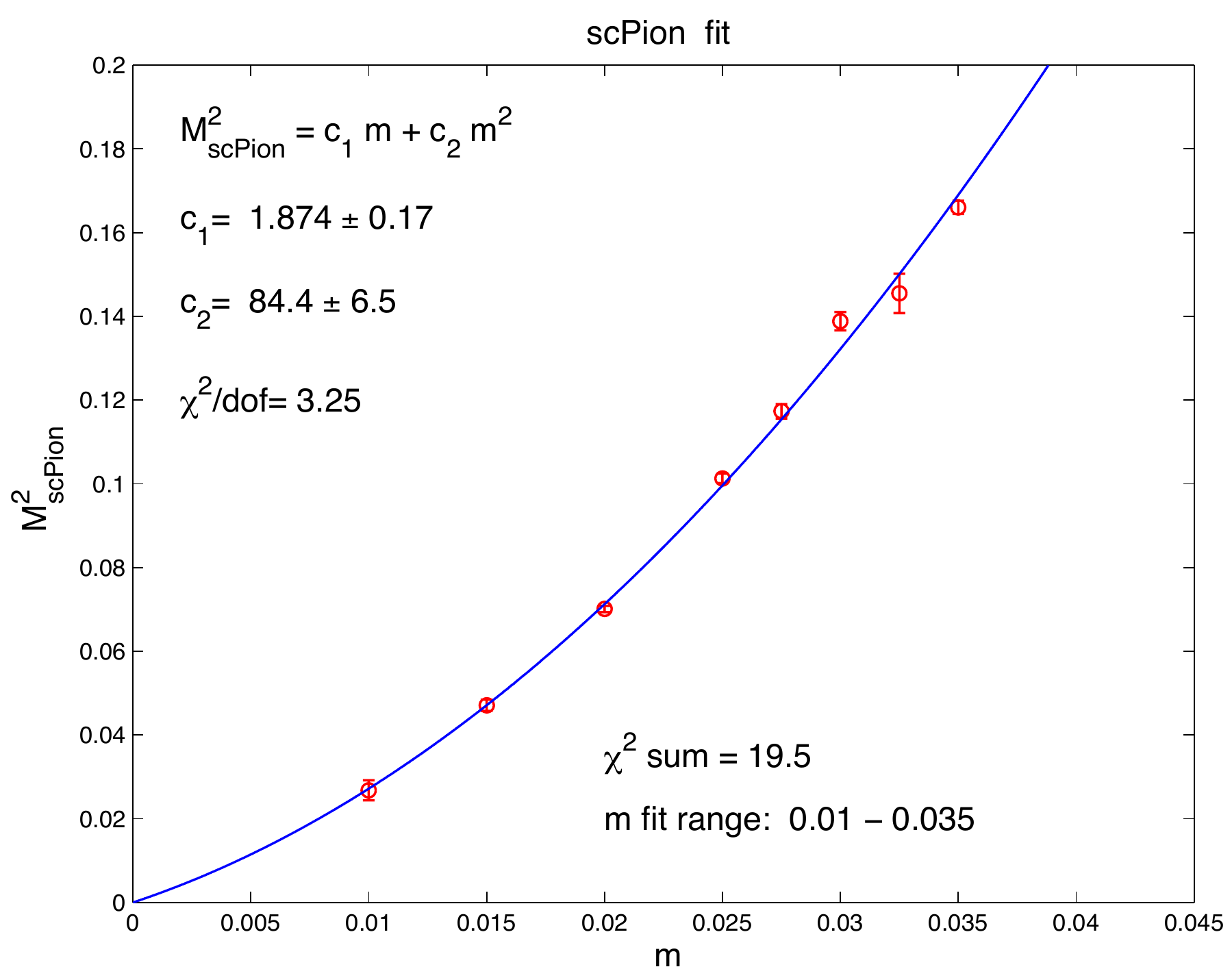}&
\includegraphics[height=4cm]{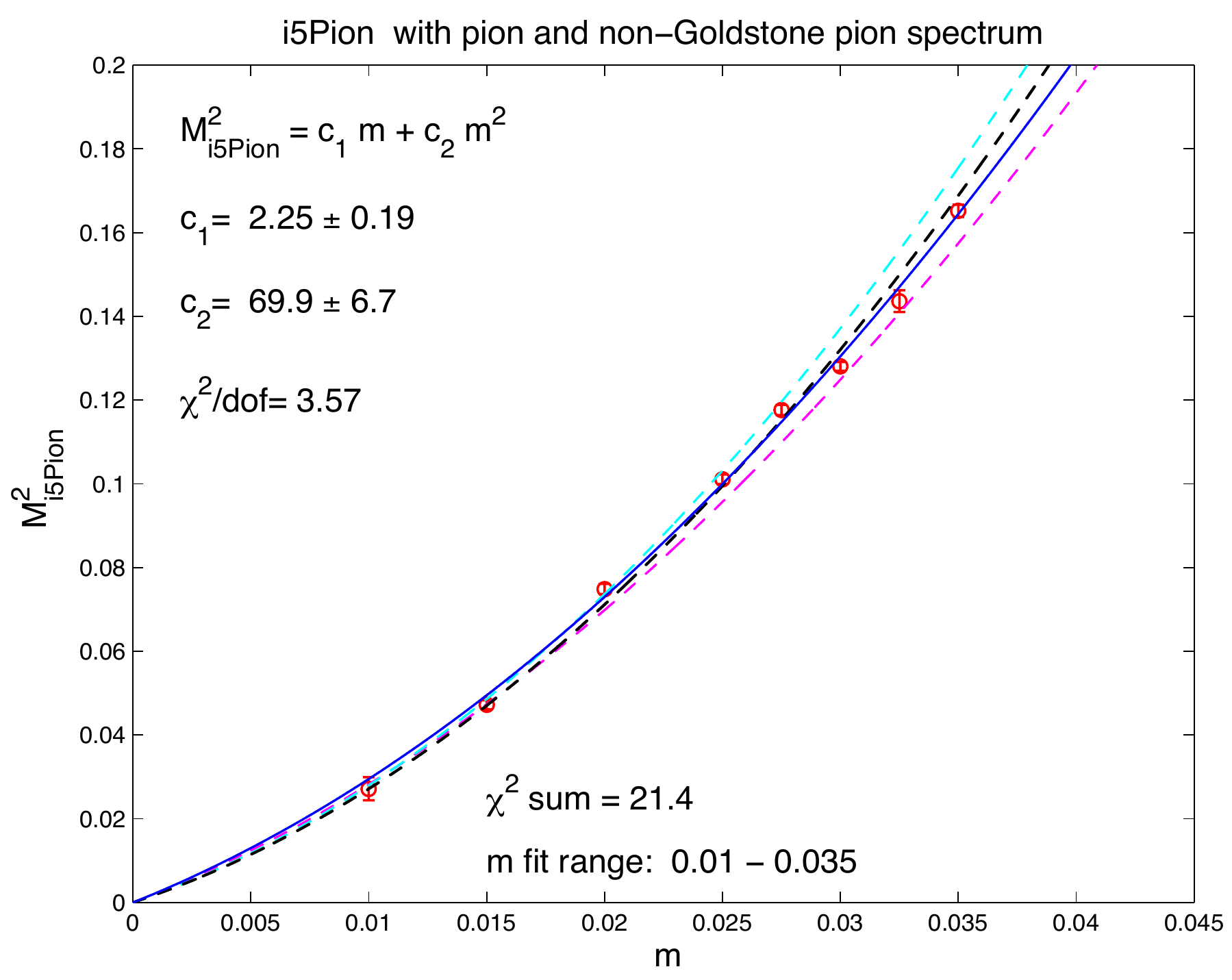}
\end{tabular}
\caption{{\footnotesize The non-Goldstone pion spectrum is shown. The composite lower right plot redisplays the i5Pion 
data together with  fits to the Goldstone pion (magenta), i5Pion (solid blue), scPion (black), and ijPion (cyan).}}
\end{center}
\label{fig:Spectrum}
\vskip -0.2in
\end{figure}

\subsection{Chiral condensate}

The chiral condensate $\langle \bar{\psi}\psi\rangle$ summed over all flavors
is dominated by the linear term in $m$ from UV contributions.  The quadratic (or linear) fit in Figure 3
gives a small non-vanishing condensate in the chiral limit which is in the expected ballpark 
from the GMOR relation  $\langle\bar{\psi}\psi\rangle=12F^2B$ with the measured low $F$  and $B$ of the order one. 
The deficit between the two sides of the GMOR relation is sensitive to the fitting procedure and  the determination of  $B$.
Adding a quadratic term to the fit is a  small effect and trying to identify chiral logs is 
beyond the scope of our simulation range. 
For an independent determination, we also studied the subtracted chiral condensate operator 
\beq
(1-m\frac{ d}{dm}|_{conn}) \cdot\langle\bar\psi\psi\rangle
\eeq
which is determined from zero momentum connected correlators.
The removal of the derivative term significantly reduces the 
dominant linear part of the $\langle \bar{\psi}\psi\rangle$ condensate. 
We find it reassuring that the two independent determinations give consistent non-vanishing results in the chiral limit
as clearly shown in Figure 3.

It should be noted that the $M_\pi$ values in the fitting range of $m$  in our analysis are {\em below}
the fitting range of  previous $N_f=12$ work on the chiral condensate work 
with considerably more uncertainty  from using the higher range~\cite{Deuzeman:2009mh}.  
In all fits we were on a fine-grained lattice in the pion mass range $aM_{\pi}=0.16-0.39$ 
and rho mass range $M_{\rho}=0.2-0.47$. In contrast, the previous study~\cite{Deuzeman:2009mh}
which reported conformal behavior was in the $aM_{\pi}=0.35-0.67$ range and rho mass range $M_{\rho}=0.39-0.77$.
Although our new results should be made even more definitive with higher accuracy and  better control on  
the systematics, the evidence is quite suggestive
for a small non-vanishing chiral condensate in the chiral limit.
\begin{figure}[t]
\begin{center}
\begin{tabular}{cc}
\includegraphics[height=4cm]{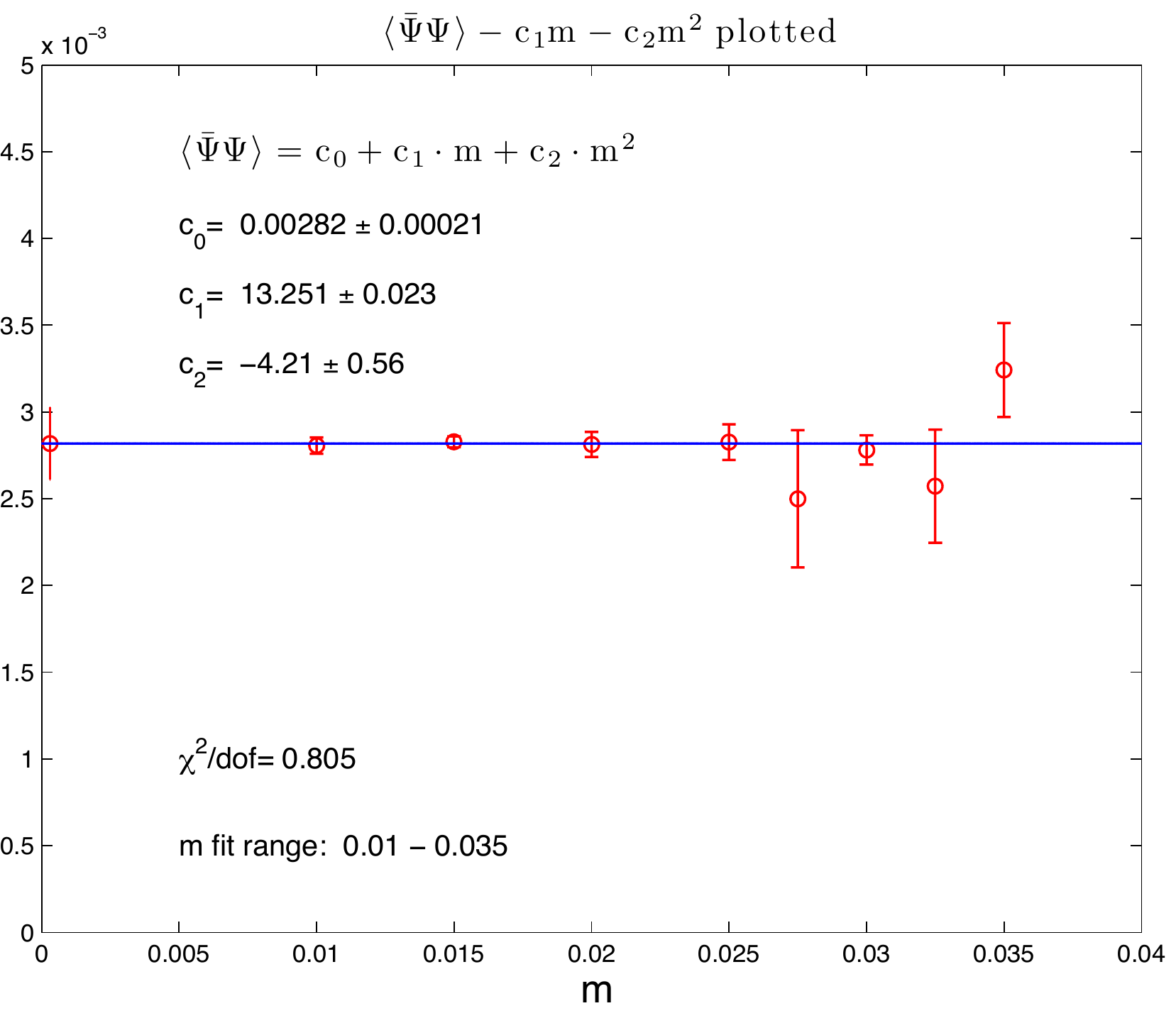}&
\includegraphics[height=4cm]{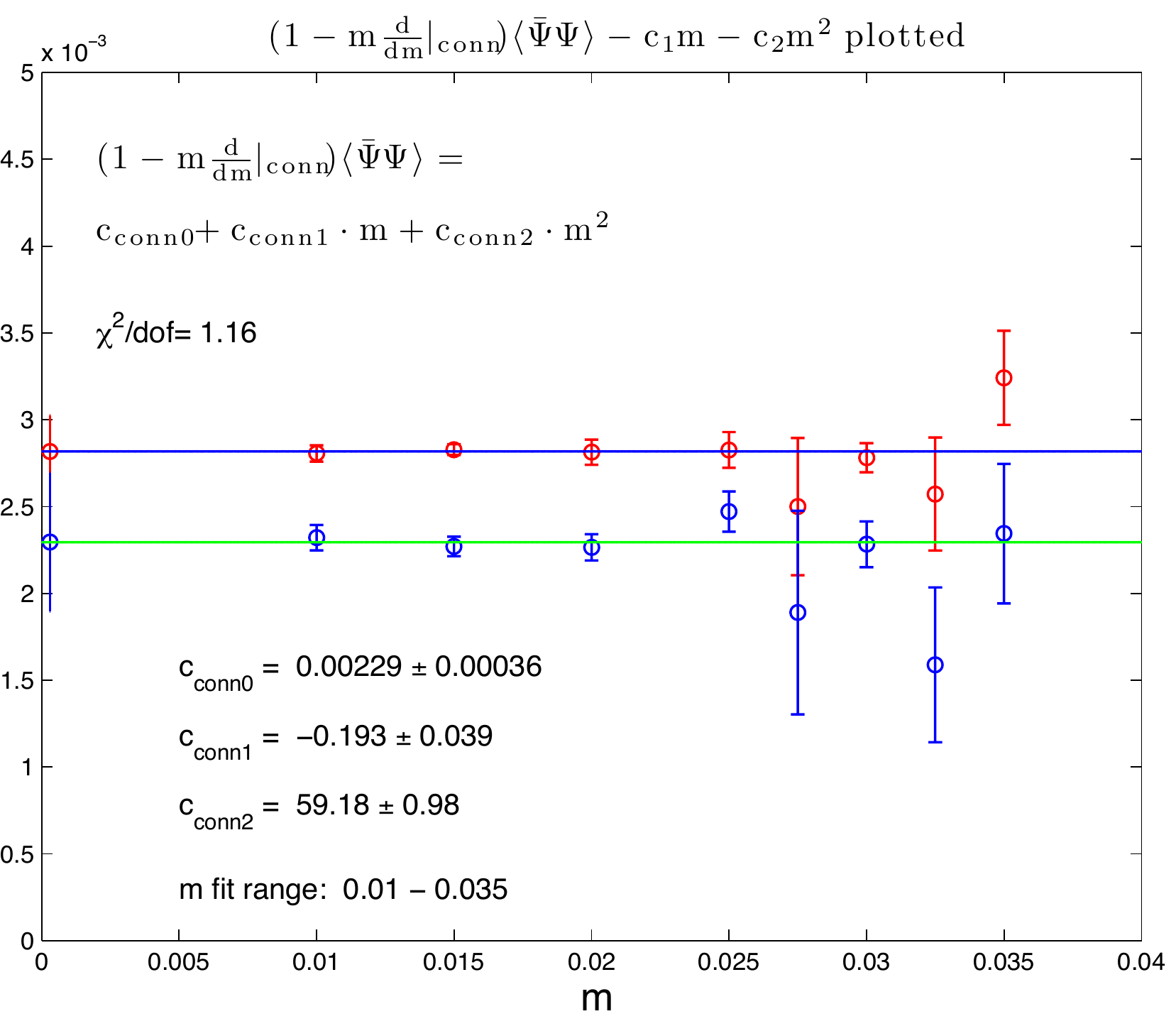}
\end{tabular}
\caption{{\footnotesize The chiral condensate and its subtracted derivative version (both have to converge  
to the same chiral limit) are shown after the removal of the non-constant part of respective fit functions of the form
$c_0+c_1m+c_2m^2$. The left side shows the fit to  $\langle \bar{\psi}\psi\rangle$ condensate data  after 
the removal of the fitted $c_1m+c_2m^2$ part with fit error on the chiral limit value of  $c_0$ at $m=0$.
The right side superimposes on the left side plot the subtracted derivative condensate fit (blue points)
(with fit coefficients displayed) after the removal 
of  the  $c_1m+c_2m^2$ part. The error of  $c_0$ at $m=0$ for this quantity is also shown and consistent
with the direct determination of the chiral limit from $\langle \bar{\psi}\psi\rangle$.
For any given $m$  always the largest volume condensate data is used since 
the finite volume analysis is not complete.
The relative sensitivity of the analysis to the lowest two or three $m$ values can be eliminated by extended systematics .}}
\end{center}
\label{fig:PbPNf12}
\vskip -0.2in
\end{figure}

\subsection{Composite hadron spectrum in the chiral limit}

\begin{figure}[h!]
\begin{center}
\begin{tabular}{cc}
\includegraphics[height=4cm]{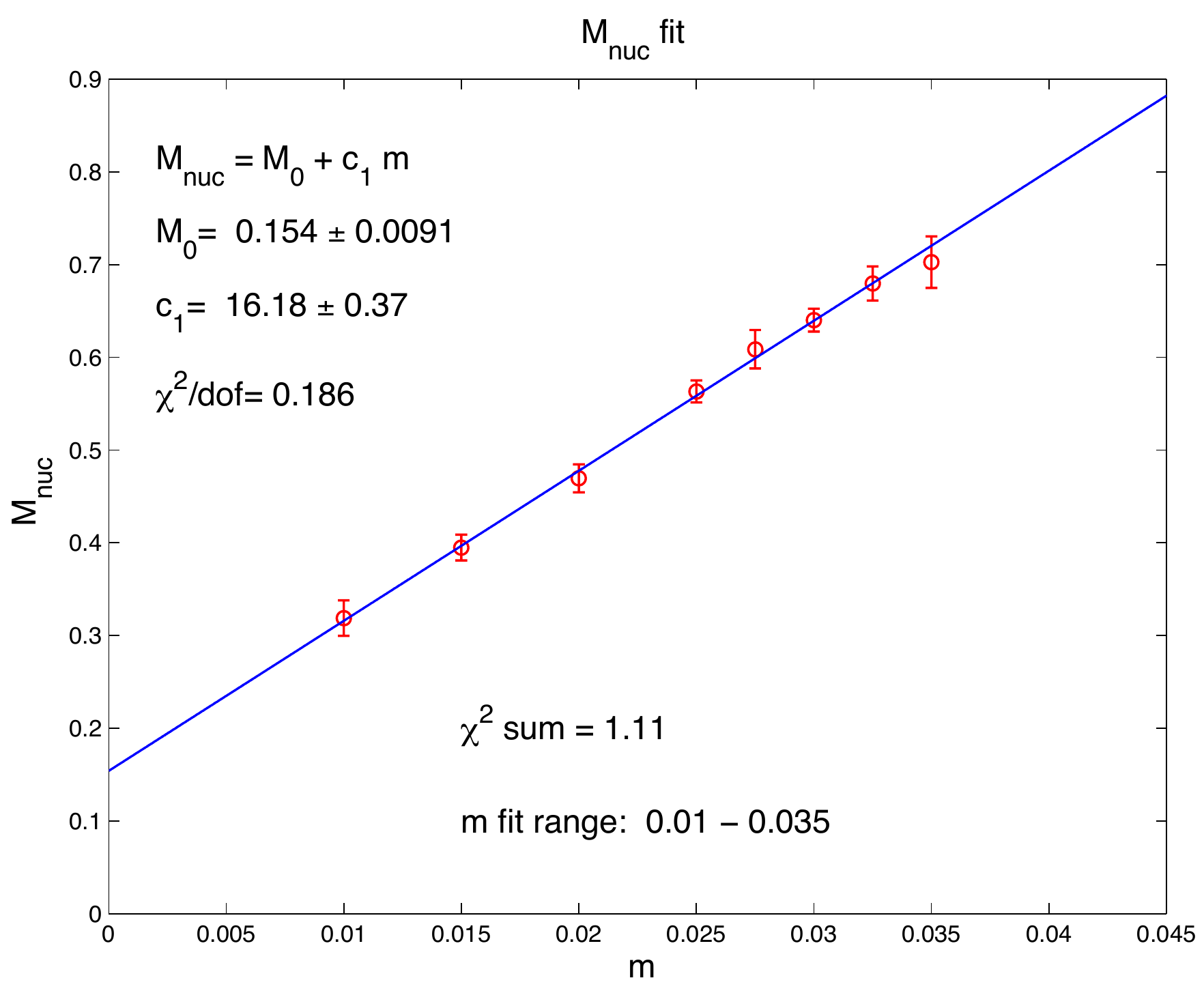}&
\includegraphics[height=4cm]{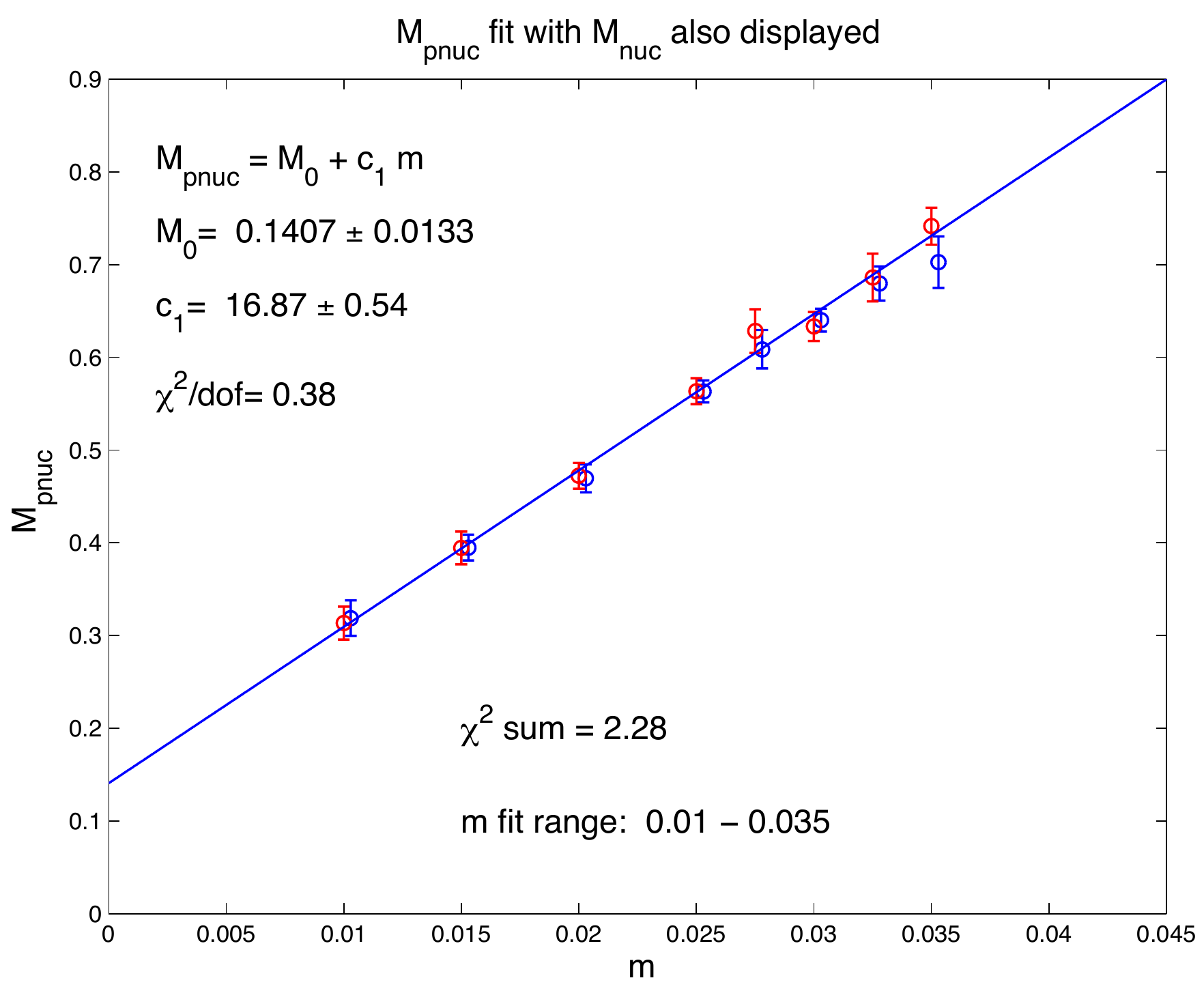}\\
\includegraphics[height=4cm]{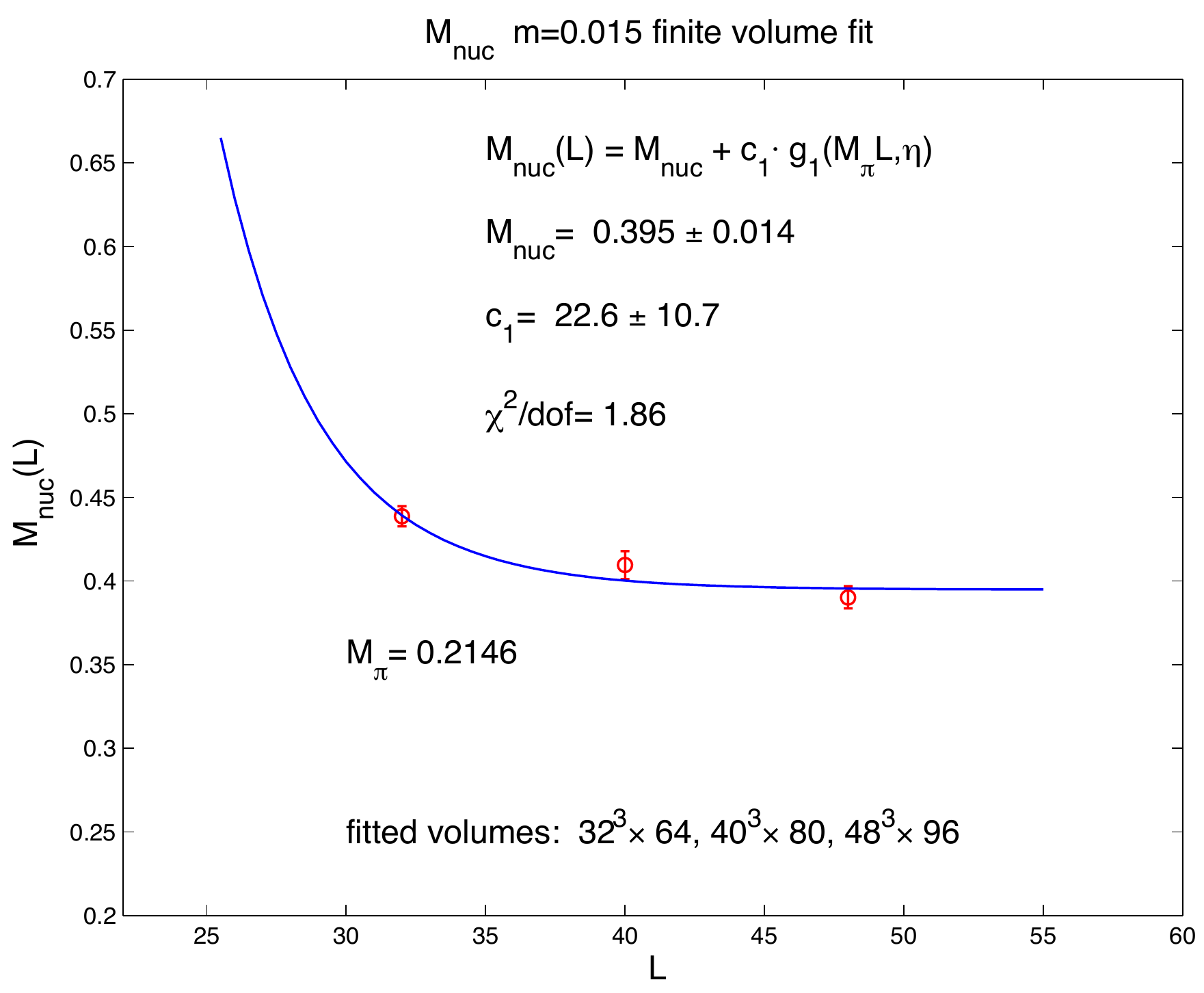}&
\includegraphics[height=4cm]{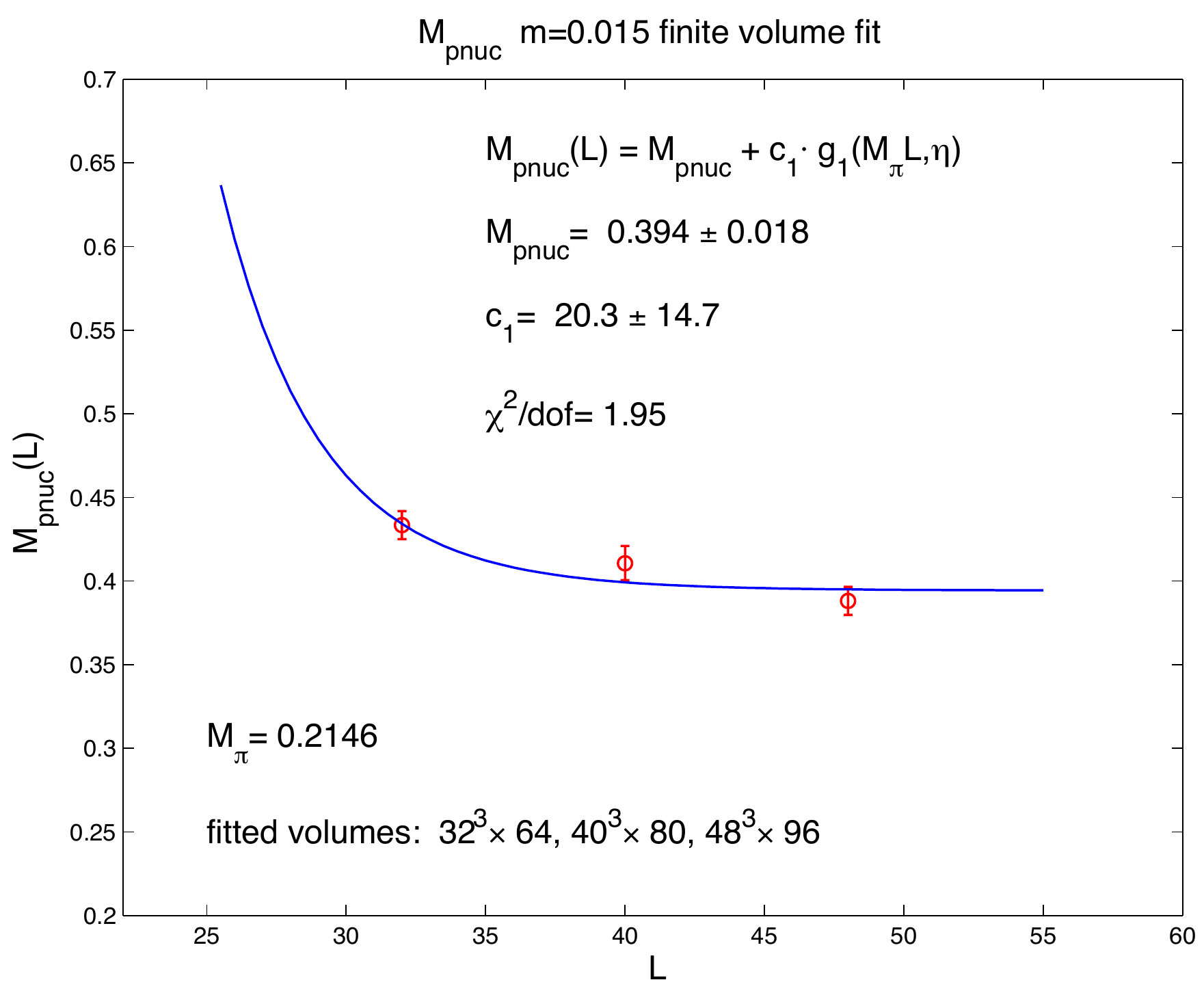}
\end{tabular}
\caption{{\footnotesize Nucleon and parity partner are fitted to the simplest linear form which is also the leading term of the chiral 
Lagrangian. The blue points on the right are the
replotted nucleon data from the left to show the degeneracy of the two states. The lower two plots show finite volume fits.}}
\end{center}
\label{fig:Nucleon}
\vskip -0.2in
\end{figure}

It is important to investigate the chiral limit of other composite hadron states. They further test the 
gaps of physical states as the fermion mass $m$ is varied and the measured hadron masses are subjected to chiral analysis
in the $m\rightarrow 0$ limit. Hadron masses also provide useful information on parity splits in several channels.
One composite state of great interest is the Higgs particle, if there is a chiral condensate close to 
the conformal window. We will briefly review new results on  the nucleon state with its parity partner, the isospin partner
of the Higgs state, and the $\rho-A_1$ splitting. 

The fermion mass dependence of the nucleon and its parity partner is shown in Figure 4 
with finite volume analysis at one selected fermion mass $m=0.015$. 
The same finite volume fit is applied as 
described earlier for the pion state. The leading chiral linear term in the fermion mass $m$ extrapolates to  non-vanishing 
chiral limit. The parity partner is practically degenerate but this is not a surprise. Already with four flavors 
a near degeneracy was reported before by the Columbia group. 

Figure 5 shows the fermion mass dependence of the Higgs particle without including the disconnected
part of the relevant correlator. Strictly speaking, therefore, the state is the $f_0$ meson with non-zero isospin.
Disconnected contributions in the correlator might shift the Higgs mass, an important issue left for future clarifications. 
Both the linear and the quadratic fits are shown together with the pseudo-goldstone
scPion which is split down from the Higgs (that is, $f_0$) state. The two states would be degenerate in the chiral limit with unbroken symmetry.
The Higgs state (as we call it) extrapolates to a nonvanishing mass in the chiral limit with  $M_H/F$ ratio
between 10 and 15. This ratio is approximately 5 in the sextet model.
\begin{figure}[htpb]
\begin{center}
\begin{tabular}{ccc}
\includegraphics[height=4cm]{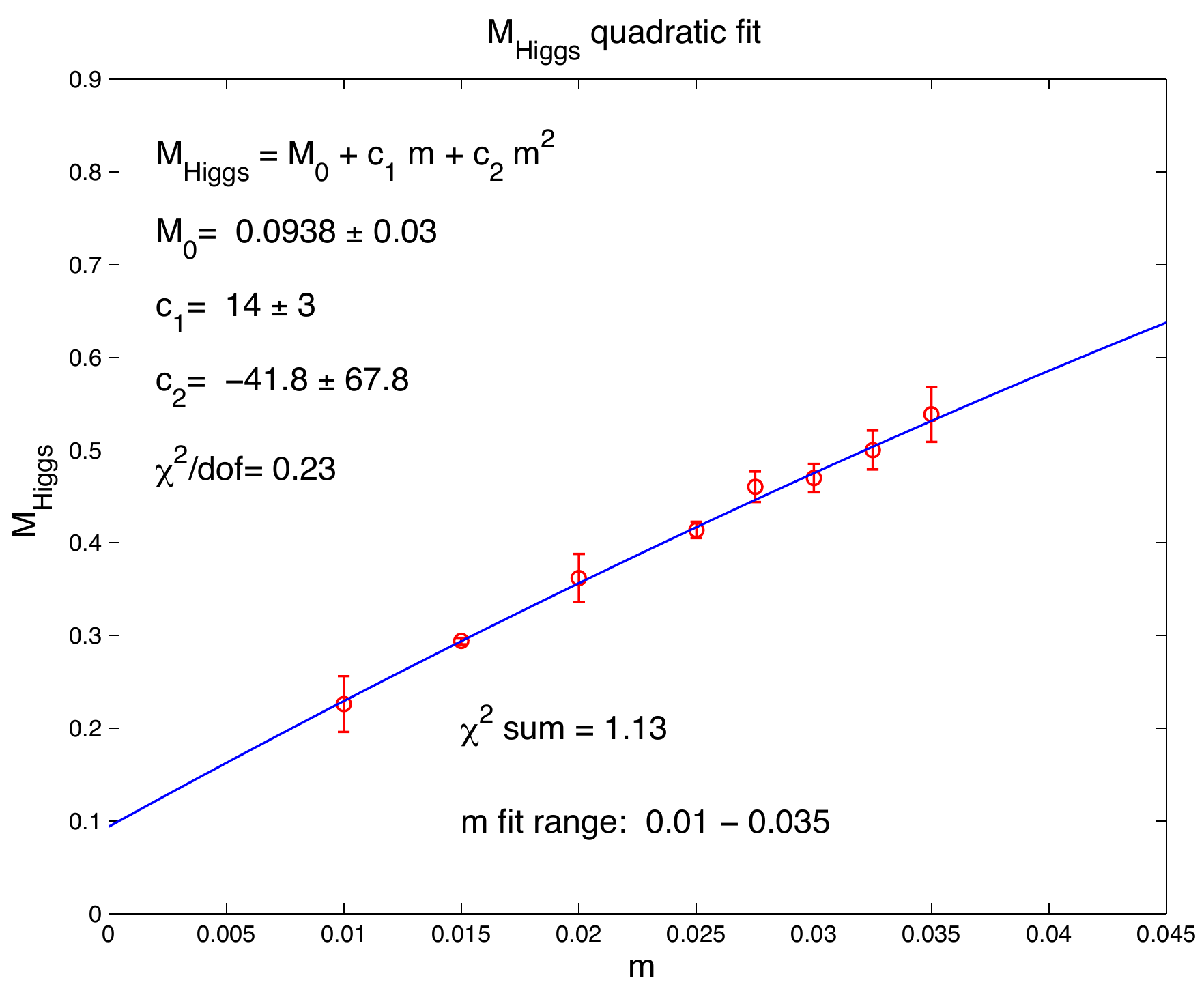}&
\includegraphics[height=4cm]{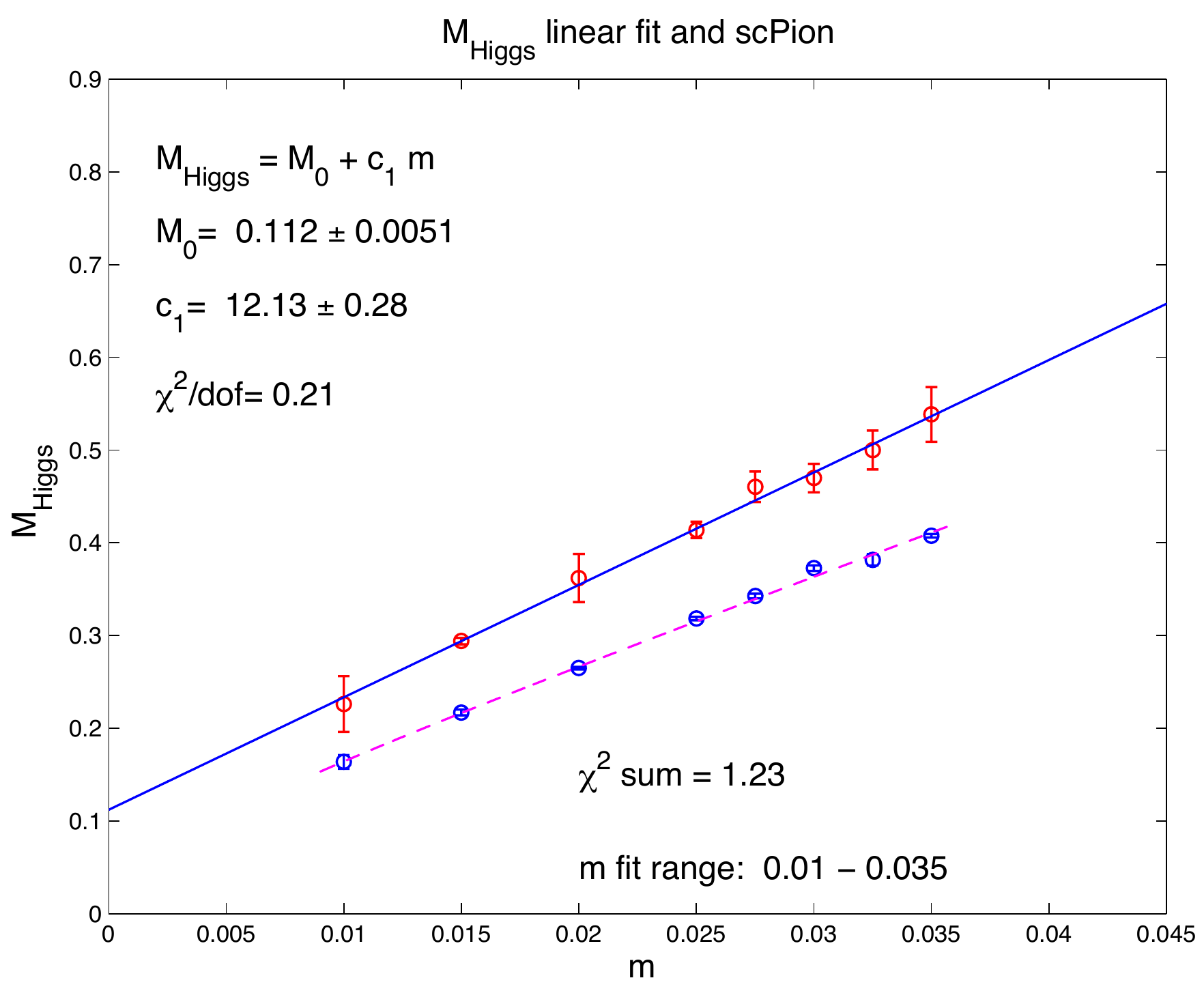}&
\includegraphics[height=4cm]{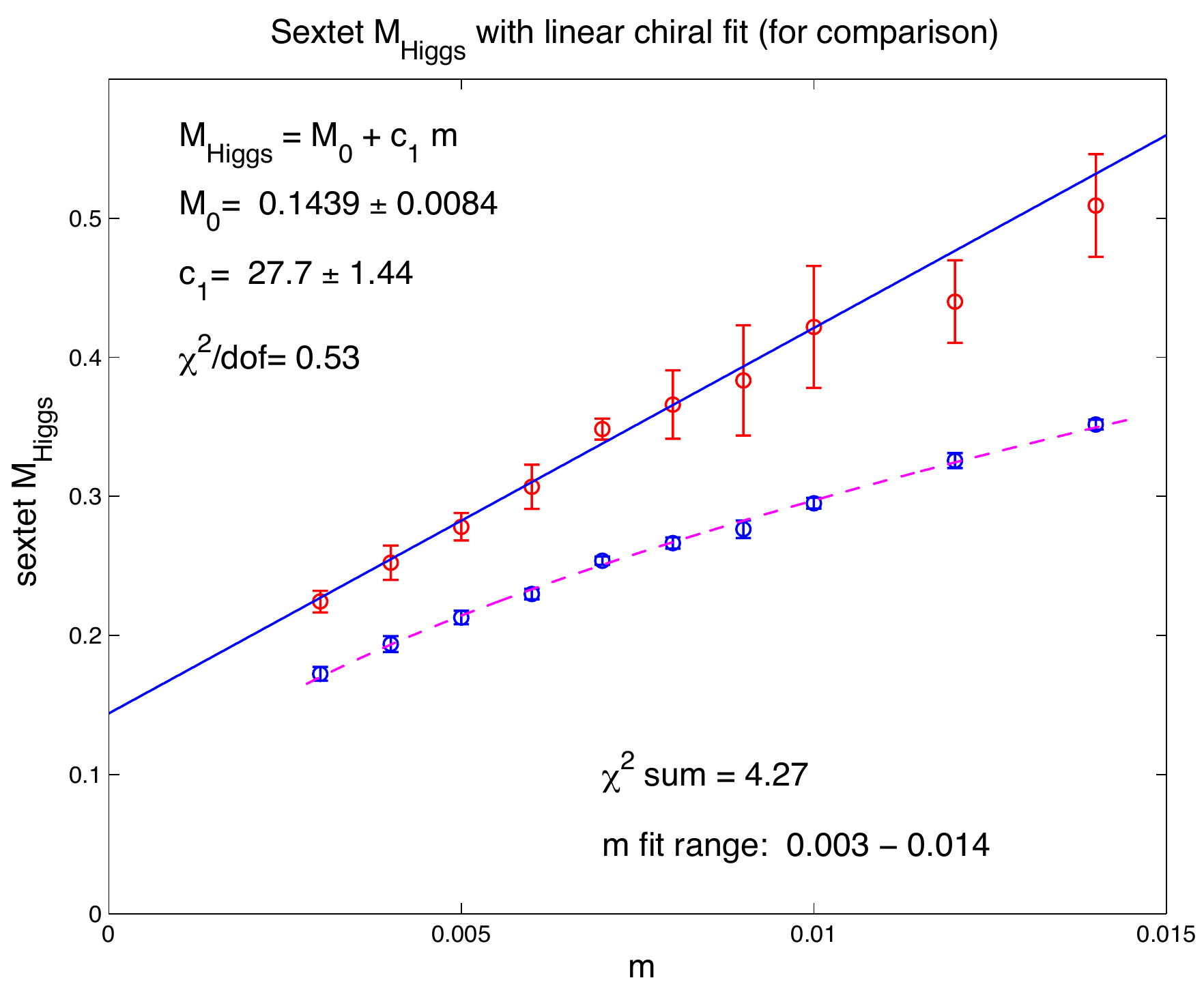}
\end{tabular}
\caption{{\footnotesize The $f_0$ state (we call it Higgs) and its splitting from the scPion state are shown. The linear fit in the middle
plot works well for the Higgs ($f_0$)
state with little change when a quadratic term is included on the left. The blue scPion
data points in the middle plot and the dashed magenta fit show the fit to the scPion state.  
The Higgs will became a resonance in the chiral limit, the missing disconnected part also contributing, so that Higgs
predictions will be challenging in future work. For comparison the Higgs state is shown on the right from the sextet analysis with its blue
scPion parity partner split with dashed magenta fit.}}
\end{center}
\label{fig:Higgs}
\vskip -0.2in
\end{figure}
Finally, Figure 6 shows the $\rho$-meson and its $A_1$ parity partner. Both states extrapolate to non-vanishing mass
in the chiral limit. The split remains significant for all fermion masses and in the chiral limit.

\begin{figure}[htpb]
\begin{center}
\begin{tabular}{cc}
\includegraphics[height=4cm]{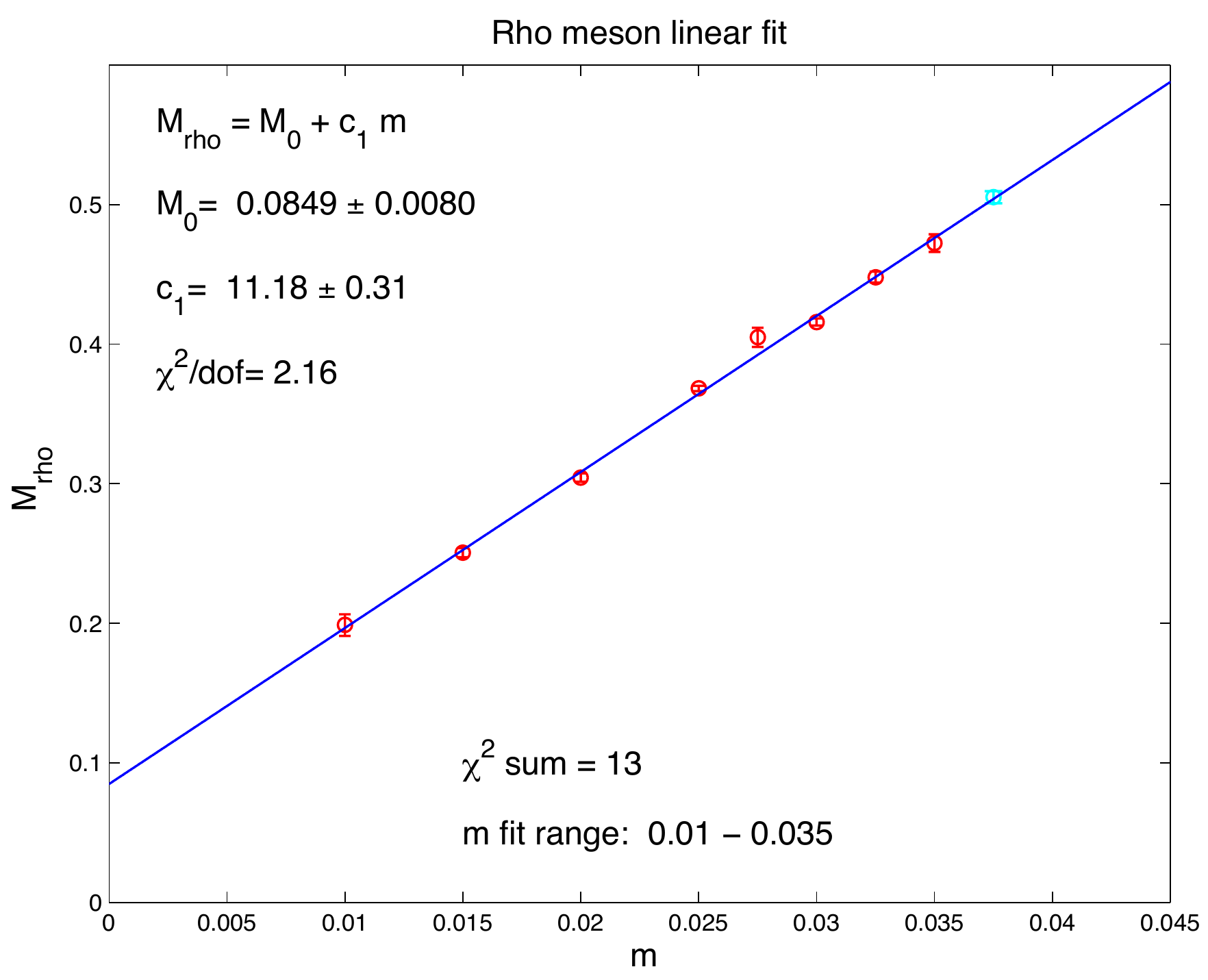}&
\includegraphics[height=4cm]{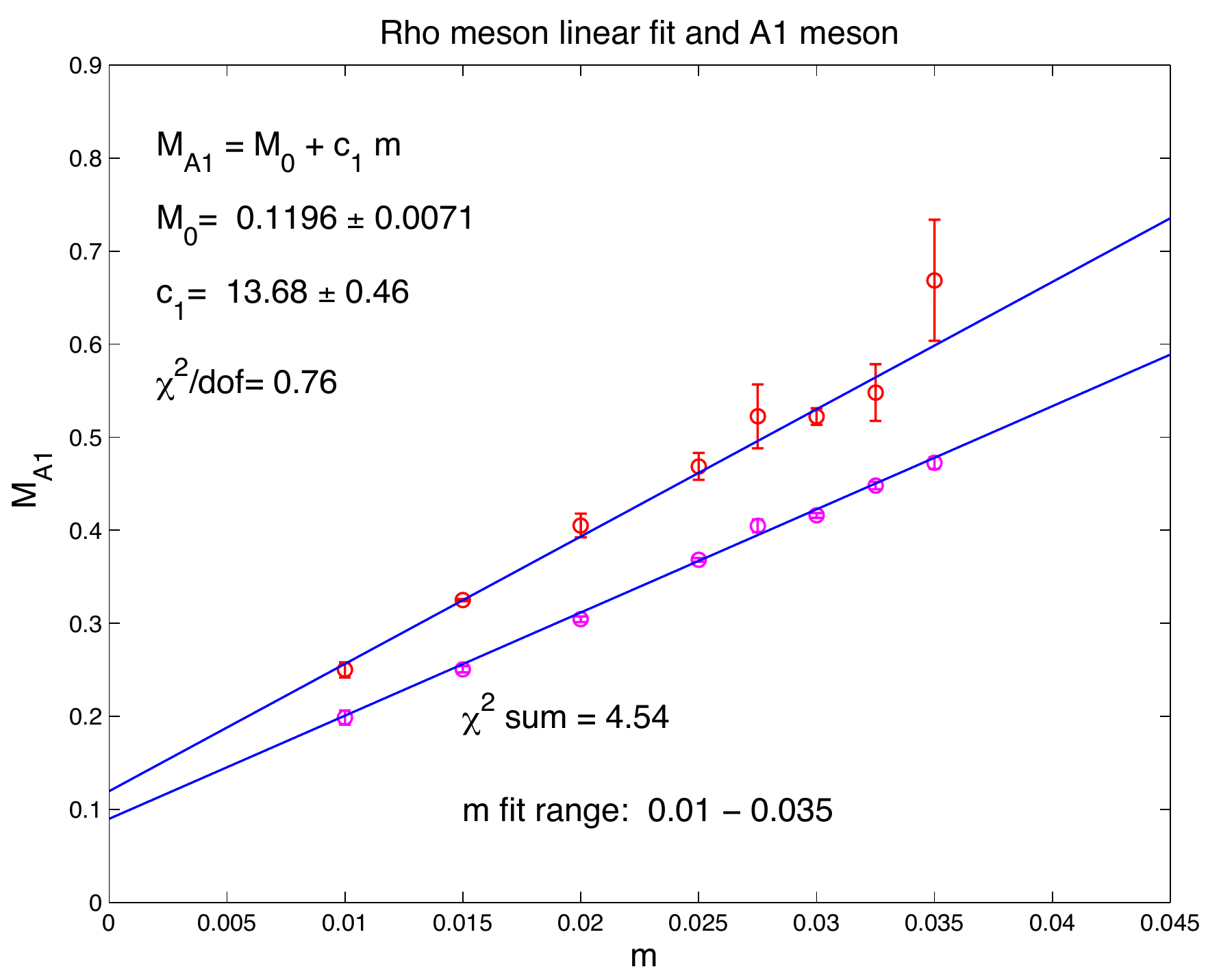}\\
\end{tabular}
\caption{{\footnotesize Rho meson and its splitting from the A1 meson are shown. 
On the right side the magenta points reproduce the data of the rho meson from the left together with its linear fit. The fit
parameters on the right show the linear fit to the $A_1$ meson.}}
\end{center}
\label{fig:Rho}
\vskip -0.1in
\end{figure}

\subsection{String tension and running coupling from the static force}

There are several ways to define a renormalized gauge coupling, for
example, the Schr\"{o}dinger Functional scheme or from square Wilson
loops. We take the renormalized coupling as defined via the
quark-antiquark potential $V(R)$, extracted from $R \times T$ Wilson
loops where the time extent $T$ can be large. From the potential, one
defines the force $F(R)$ and coupling $\alpha_{qq}(R)$ as
\be
F(R) = \frac{dV}{dR} = C_F \frac{\alpha_{qq}(R)}{R^2}, \hspace{0.4cm}
\alpha_{qq}(R) = \frac{g^2_{qq}(R)}{4 \pi}.
\label{eq:F}
\ee
The coupling is defined at the scale $R$ of the quark-antiquark
separation, in the infinite-volume limit $L \rightarrow \infty$. This
is different from the scheme using square Wilson loops, where one has
$\alpha_W(R,L)$ and one can choose finite $R$ with $L \rightarrow \infty$,
or finite $L$ and fixed $R/L$ ratio. In the former case, these schemes
are related via 
\be
\alpha_{qq}(R) = \alpha_W(R) [ 1 + 0.31551 \alpha_W(R) + {\cal
  O}(\alpha_W(R)^2) ].
\label{eq:qqw}
\ee
The $\beta$ function in the qq scheme is known to 3-loops.
For $SU(3)$ gauge theory with $N_f=12$ fundamental
flavors, the location of the infrared fixed point to 3-loop
order is $\alpha_{qq}^* = 0.3714...$ This is about 50\% of the
scheme-independent 2-loop value of $\alpha^*$, indicating that
higher order corrections beyond 3-loop might not be negligible.
\begin{figure}[htpb]
\begin{center}
\includegraphics[width=0.3\textwidth]{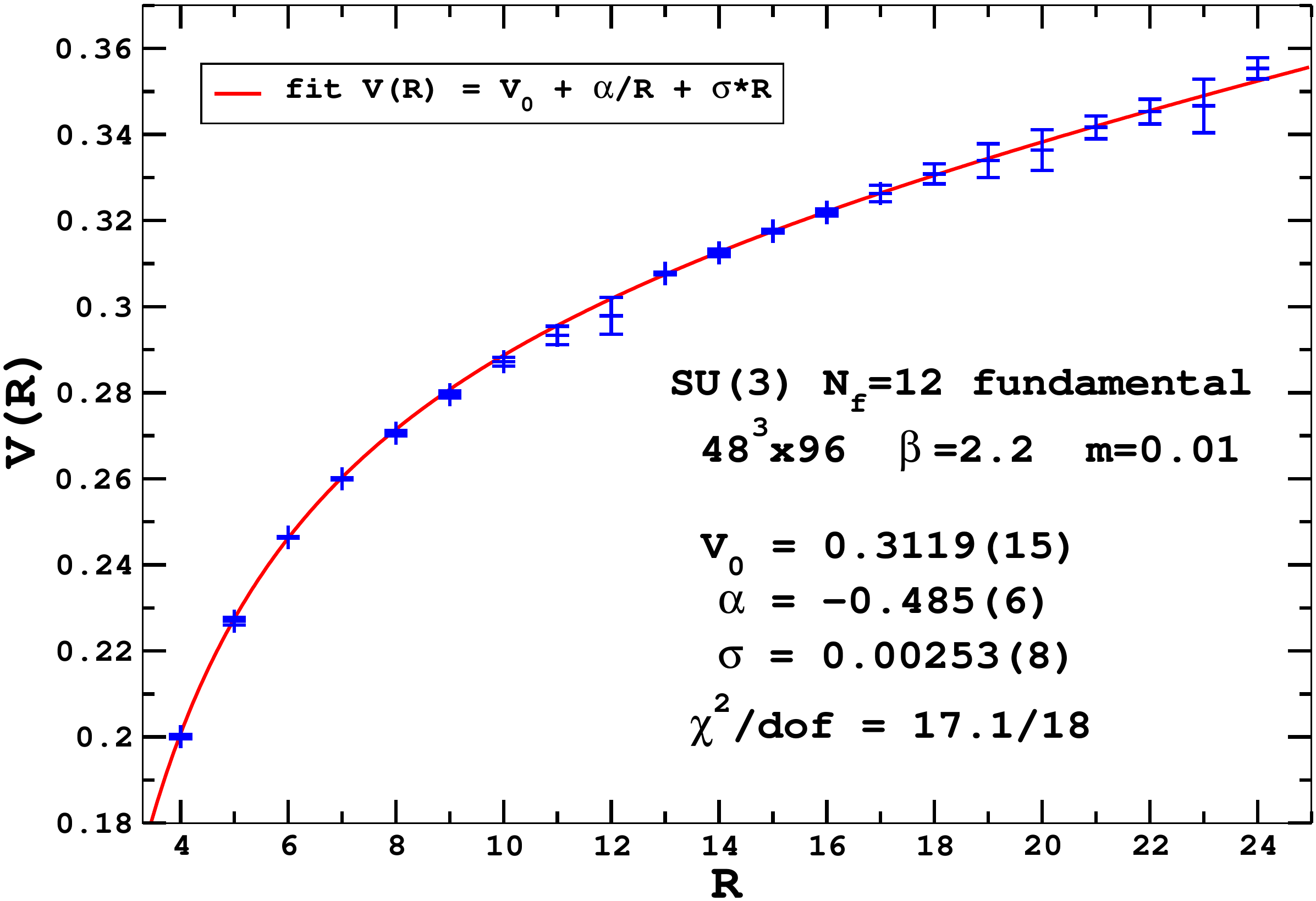}
\hspace{0.4cm}
\includegraphics[width=0.3\textwidth]{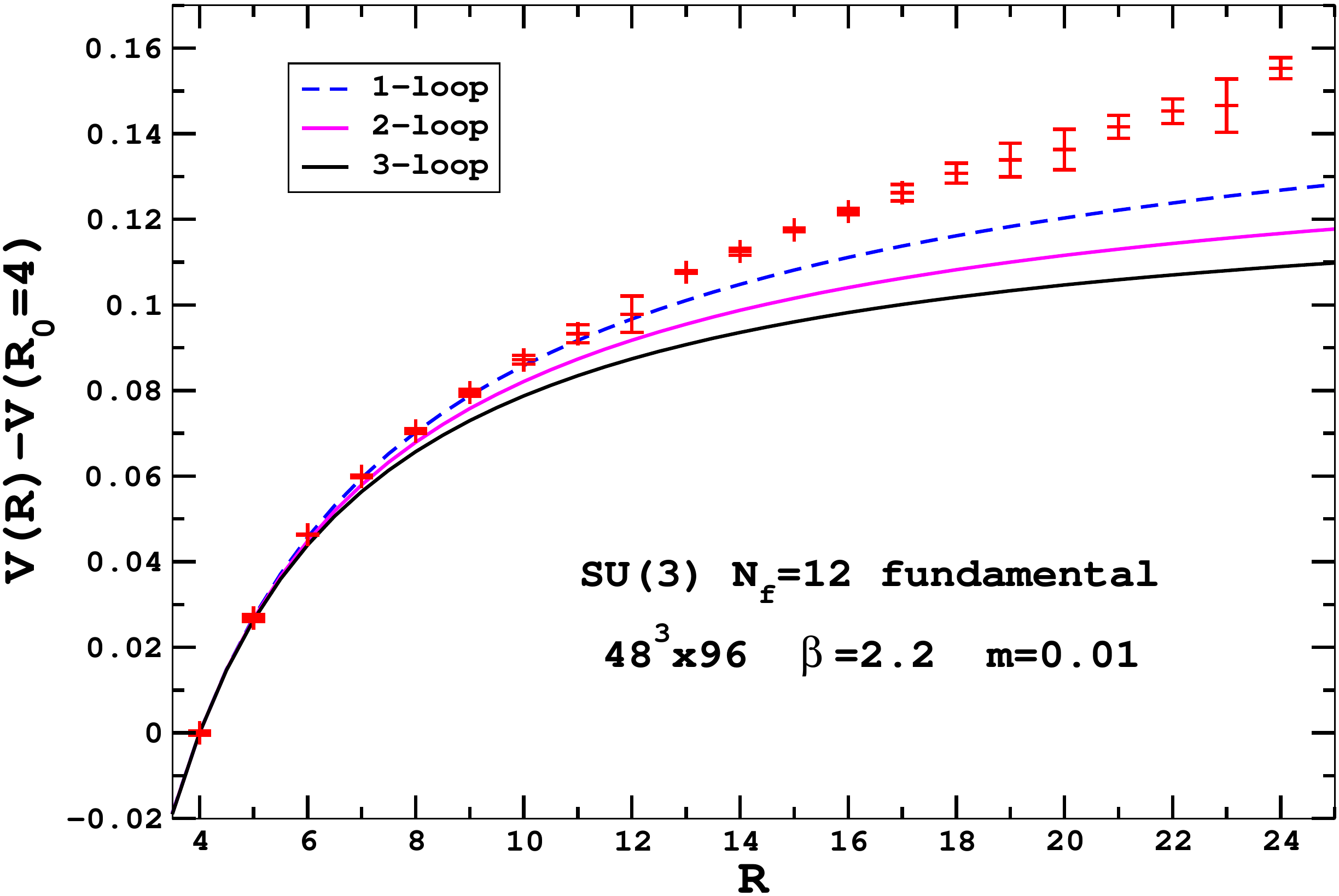} 
\hspace{0.4cm}
\includegraphics[width=0.3\textwidth]{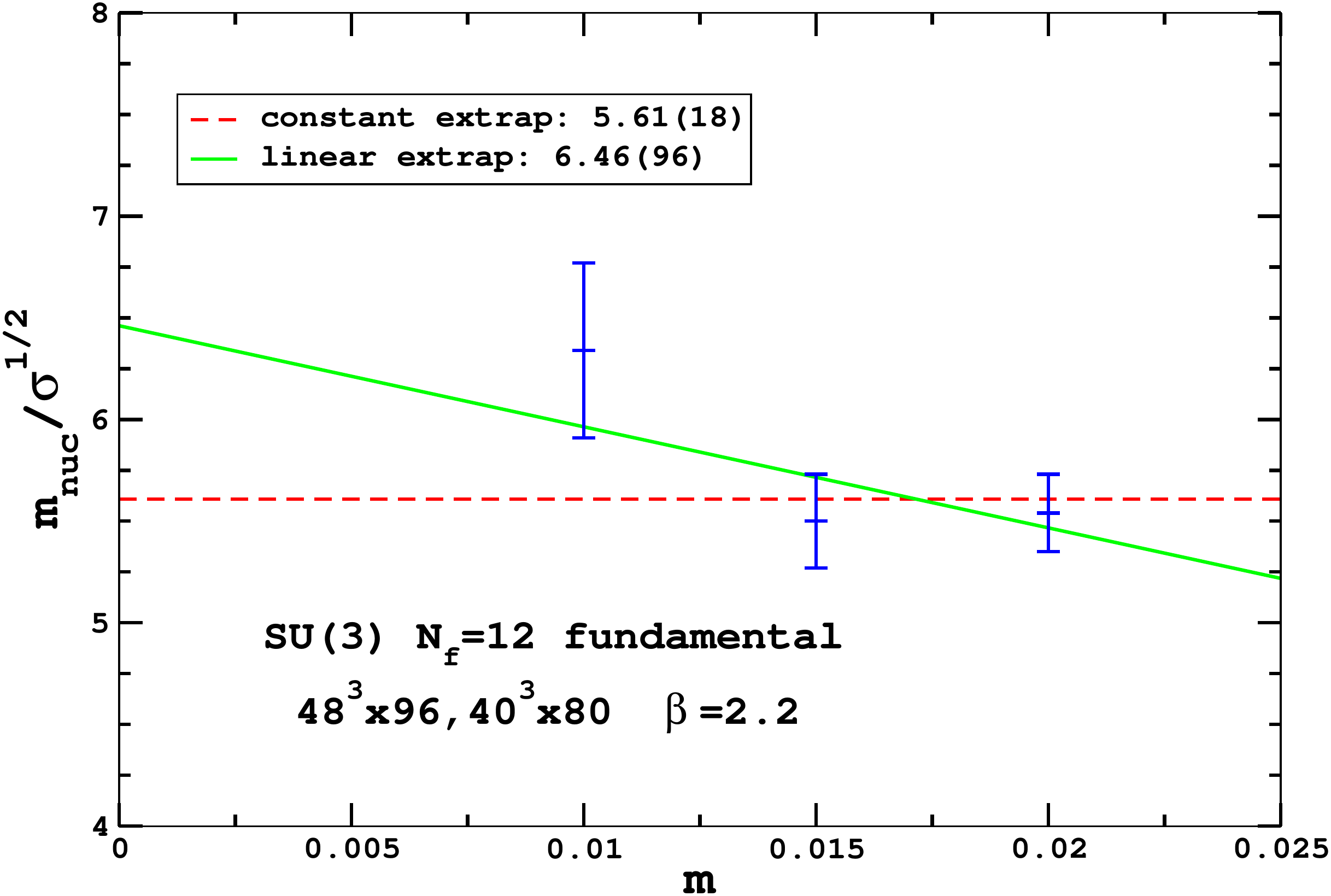} 
\caption{{\footnotesize $V(R)$ data and fit for  $m=0.01$ is plotted on the left and comparison with
perturbation theory is plotted in the middle. The right side plot shows
the string tension  measured in nucleon mass units at $m=0.01,0.015,0.02$ and extrapolated to the chiral limit. 
The finite nucleon mass  gap in the chiral limit implies finite  string tension at $m=0$. }}
\label{fig2}
\end{center}
\end{figure} 

 A range of lattice spacings, volumes and quark masses are studied in the running coupling project, we
show results for the largest volume $48^3 \times 96$ at $\beta=2.2$
and quark masses $m=0.01$ and 0.015 and for the $40^3 \times 80$ run at $m=0.02$.
To improve the measurement of $V(R)$, we use different levels
of APE-smearing to produce a correlation matrix of Wilson loops, the
lowest energy is extracted using the generalized eigenvalue method. We
also improve the lattice force, which is naively discretized as
$F(R+1/2) = V(R+1) - V(R)$. For the Symanzik gauge action, the
improvement is a relatively small effect, for example the naive value
$R+1/2 = 4.5$ is shifted to $4.457866...$

In Figure 7 on the left we show the measured $V(R)$ fitted to 
the form
\be
V(R) = V_0 + \frac{\alpha}{R} + \sigma R.
\ee
for $m=0.01$. The $m=0.015$ and $m=0.02$ runs are shown on the right of Figure 7. 
For all three masses, the resulting fits are good, with a clear signal of
linear dependence and an effective string tension $\sigma$. The string
tension decreases with the quark mass, its behavior in conjunction
with the mass spectrum in the chiral limit is under investigation and the first result is shown in the figure.
The finite nucleon mass  gap in the chiral limit implies finite  string tension at $m=0$.

The renormalized coupling $\alpha_{qq}(R)$ is a derivative of the
potential $V(R)$ and hence more difficult to numerically measure via
simulations. The most accurate comparison between lattice simulations
and perturbation theory is directly of the potential $V(R)$
itself. This is naturally given by finite potential differences
\be
V(R) - V(R_0) = C_F \int_{R_0}^R \frac{\alpha_{qq}(R')}{R'^2} dR',
\ee
where $R_0$ is some reference point where $\alpha_{qq}(R_0)$ is
accurately measured in simulations. From this starting point, the
renormalized coupling runs according to perturbation theory, at some
loop order. The result is shown in the middle of Figure 7, with curves at
1-, 2- and 3-loop order for the potential difference. 
Although progress was made in studies of important finite volume effects, more work is needed to bring the systematics
under full control.
In the current state of the analysis the string tension and the fast running coupling
are consistent with the $\chi SB$ hypothesis and do not support the conformal one.

\subsection{Testing the alternate hypothesis of conformal chiral symmetry}

\begin{figure}[htpb]
\begin{center}
\begin{tabular}{cc}
\includegraphics[height=4cm]{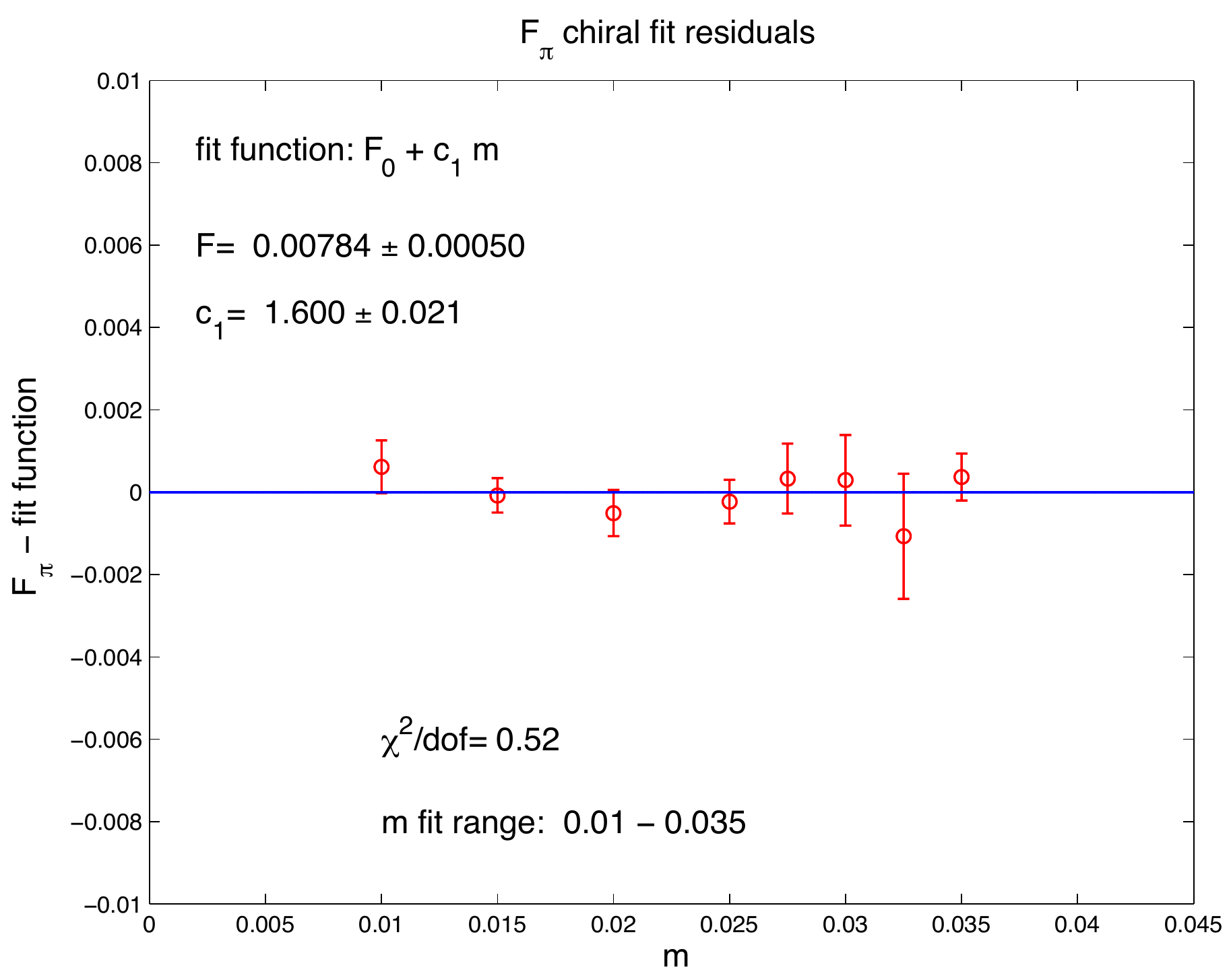}&
\includegraphics[height=4cm]{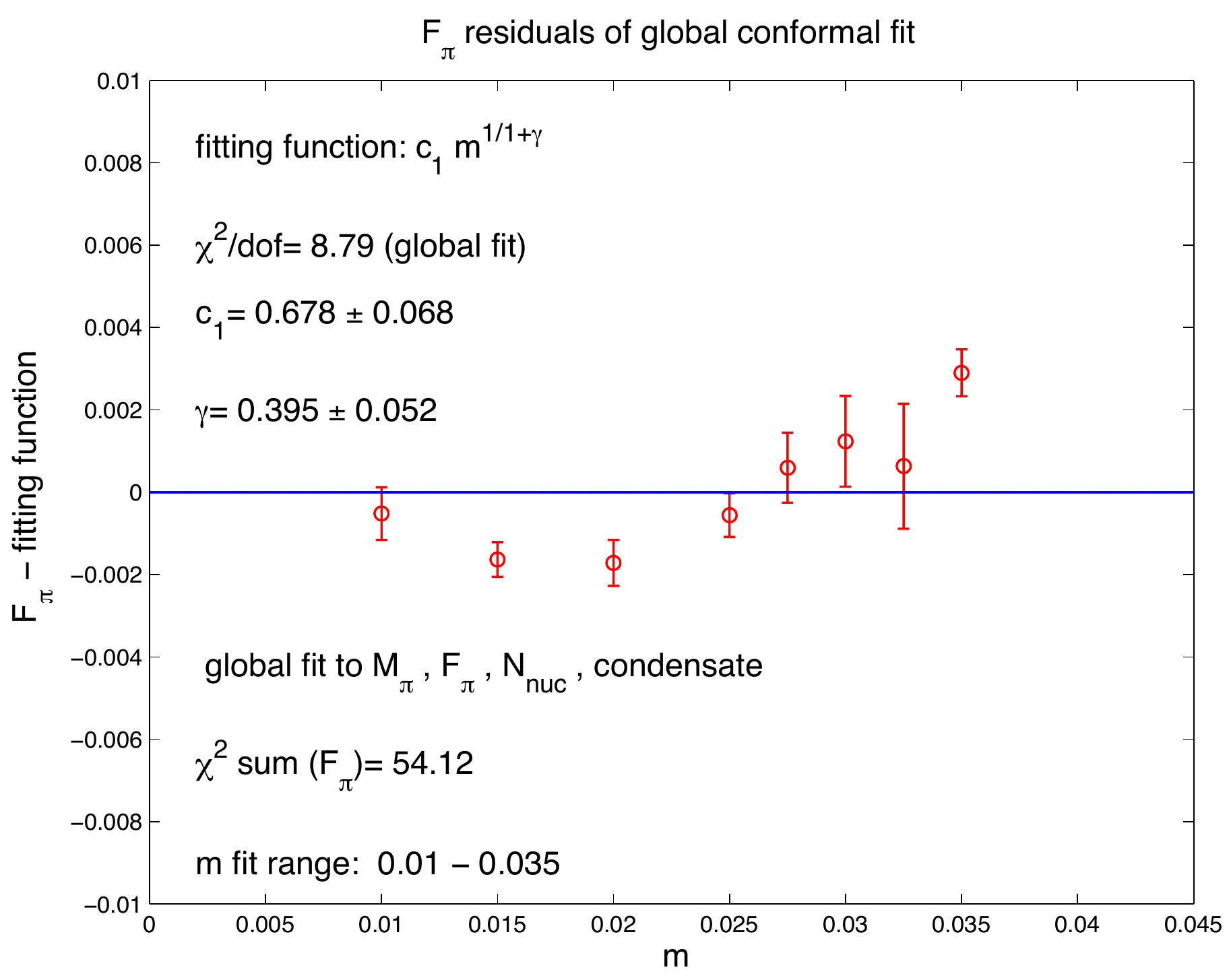}\\
\includegraphics[height=4cm]{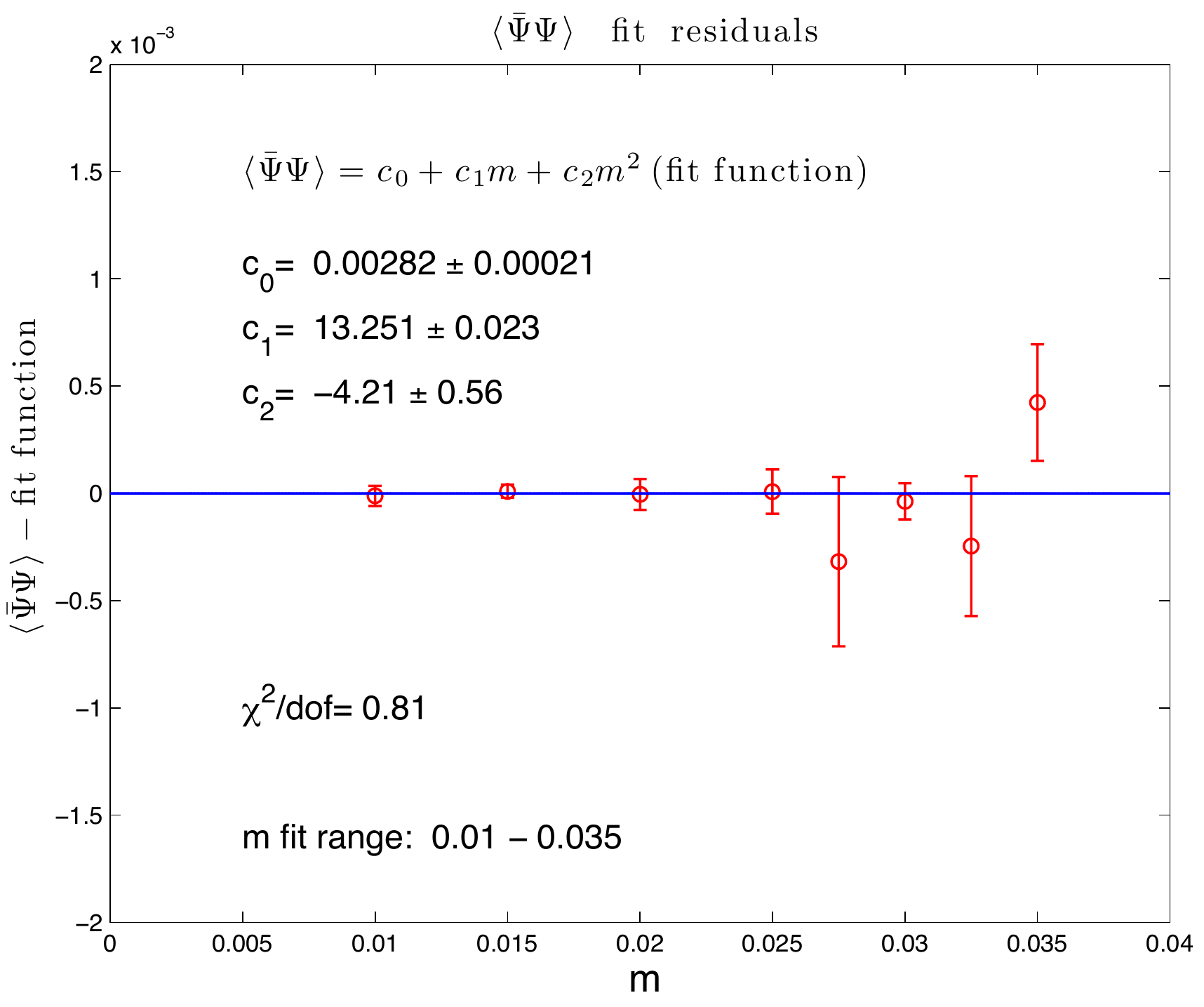}&
\includegraphics[height=4cm]{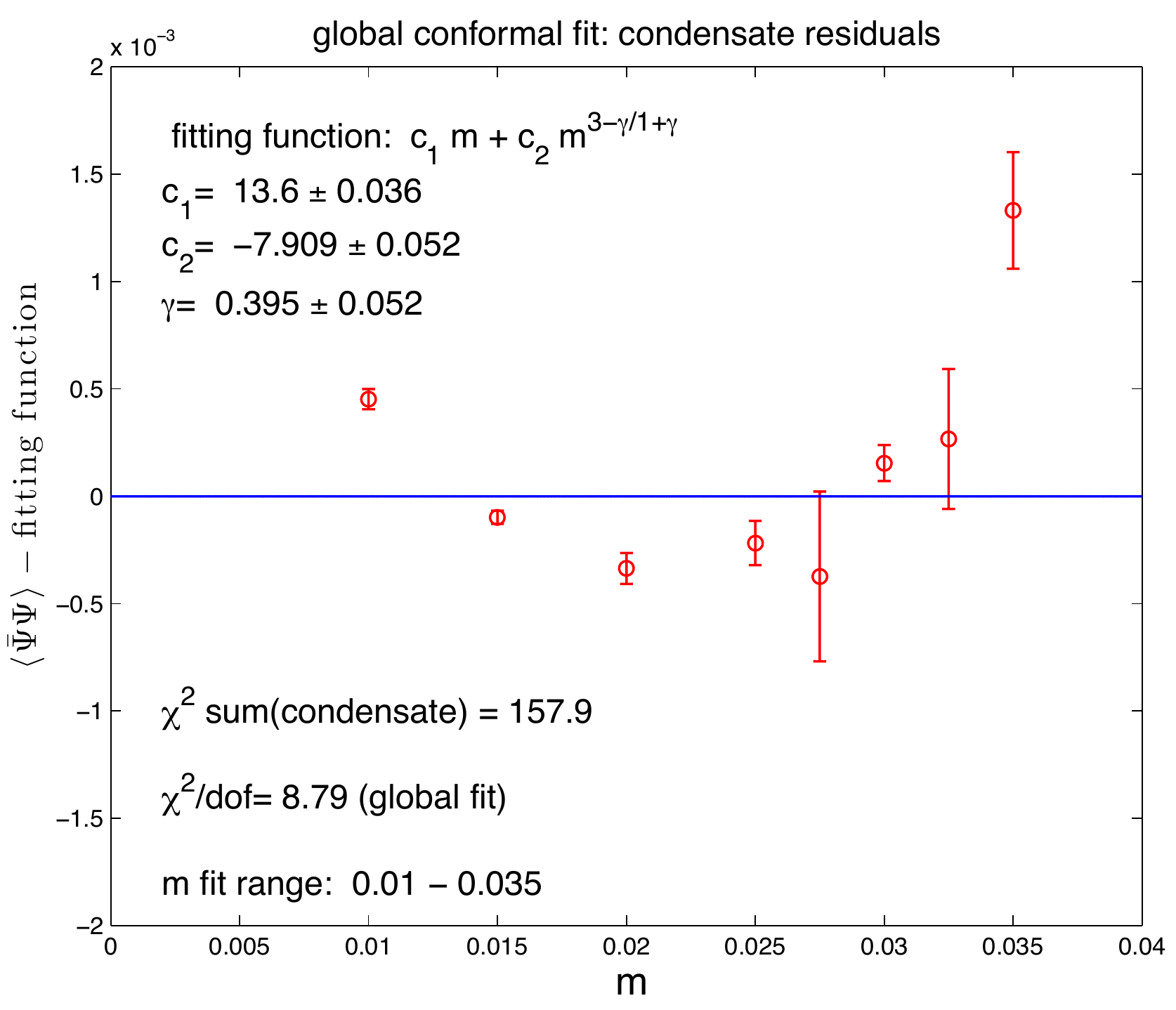}
\end{tabular}
\caption{{\footnotesize The $N_f=12$ chiral and conformal simultaneous fits in four channels are displayed for 
comparison in two select cases.}}
\end{center}
\label{fig:ConformNf12}
\vskip -0.2in
\end{figure}

The simulation results we presented  for twelve fermions in the fundamental
repesentation of the $SU(3)$ color gauge group  favor the chiral symmetry
breaking hypothesis. The pion state is consistent with a vanishing mass in the chiral limit
and easy to fit with a simple quadratic function of the fermion mass. The non-Goldstone pion spectrum 
shows very little taste breaking at $\beta=2.2$ and the small splittings are consistent with expectations
for staggered fermions with stout smearing. The $SO(4)$ degeneracies and splittings appear to follow the
pattern of QCD although the fermion mass dependence is significantly different. The fundamental 
scale-setting parameter $F$ of chiral symmetry breaking is finite in the chiral limit.
A non-vanishing chiral condensate is found in the chiral limit which is in the expected ballpark of the GMOR relation as 
suggested by the small value of $F$. The subtracted chiral condensate, after the dominant linear UV-contribution
is removed, also yields a consistent non-vanishing condensate in the chiral limit. The nucleon states, 
the Higgs ($f_0$) meson, the $\rho$ meson and $A_1$ meson extrapolate to non-vanishing masses in 
the chiral limit and considerable splits of some of the parity partner states persist at very low fermion masses toward the
chiral limit. There seems to be an effective string tension indicating confinement-like behavior below the string-breaking scale
and the running coupling has 
not shown  signs of a fixed point slowdown.  In addition, there seems to be a rapid finite
temperature transition whose nature is unclear but hardly favors a conformal bulk phase. 
Our results are consistent with results reported in ~\cite{Jin:2009mc} but disagree with the chiral analysis
of ~\cite{Deuzeman:2009mh} and do not support the infrared fixed point reported
in~\cite{Appelquist:2007hu}.

But is it possible that we mislead ourselves with the $\chi{\rm SB}$ interpretation? Can we interpret the results as
conformal chiral symmetry? To decide this question, a fairly stringent test is possible. With the conformal hypothesis
the mass dependence 
of all physical states is controlled by the anomalous dimension $\gamma$  for small fermion masses~\cite{DelDebbio:2010hx}. Each
hadron and $F_\pi$ should scale as  $M_\pi \approx m^{1/y_m}$ and $F_\pi \approx m^{1/y_m}$ for small $m$ where 
$y_m=1+\gamma$. For small enough $m$ the value of $\gamma$ should be interpreted 
as $\gamma^{*}$ at the infrared fixed point. The chiral condensate is expected to have the behavior 
$\langle \bar{\psi}\psi\rangle \approx c\cdot m +  m^{\frac{3-\gamma}{1+\gamma}}$ when $m\rightarrow 0$.
We selected various subsets of states for a combined fit with universal critical exponent $\gamma$. We also fitted all measured
states combined. 
Applying the
conformal hypothesis to the chiral condensate, to $F$, to the pion state, and to the stable nucleon state collectively leads to
a results with a $\chi^2$ sum of  $\chi^2=229$ for 26 degrees of freedom with  $\chi^2/{\rm dof}=8.79$. This indicates very low 
level of confidence in the hypothesis. The  $\chi{\rm SB}$ hypothesis gives  $\chi^2/{\rm dof}=1.22$ for the same set of states. 
This was the result quoted in Section 1. 
The chiral and conformal fits for two of the four fitted states with the quoted global results is shown in Figure 8.
Applying the global analysis to all states we measured, the contrasting behavior is less pronounced but still significant. 
The results disfavoring the conformal hypothesis are significant. More work is needed for higher accuracy and full control of the systematics.

\section{Two fermions in sextet representation of the SU(3) color gauge group}

 Our findings in the sextet model are consistent with chiral symmetry
breaking.
% but finite volume studies, simulations at several gauge couplings to vary the lattice scale, to study
%the running coupling and the finite temperature phase transition all require new and significant resources.
The gauge coupling $\beta=3.2$ where we report new results is more coarse-grained than the $\beta=2.2$ set 
in the $N_f=12$ model. Simulations at weaker bare coupling are underway. 
The chiral and conformal fitting procedures are identical to those we described earlier in the $N_f=12$ fermion model.

\subsection{Sextet Goldstone spectrum and $\bf F_\pi$ from  chiral symmetry breaking}
\begin{figure}[!htpb]
\begin{center}
\begin{tabular}{cc}
\includegraphics[height=4cm]{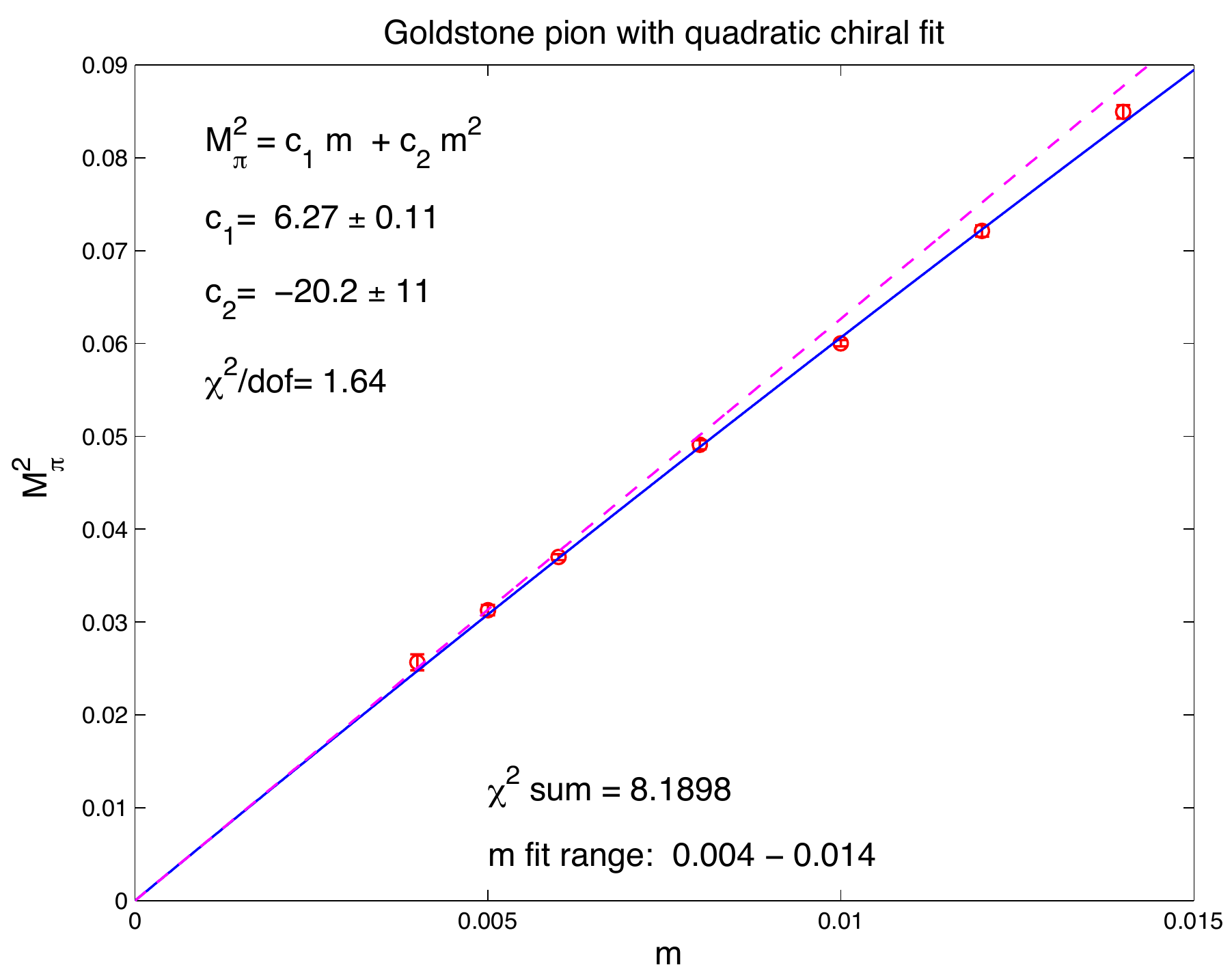}&
\includegraphics[height=4cm]{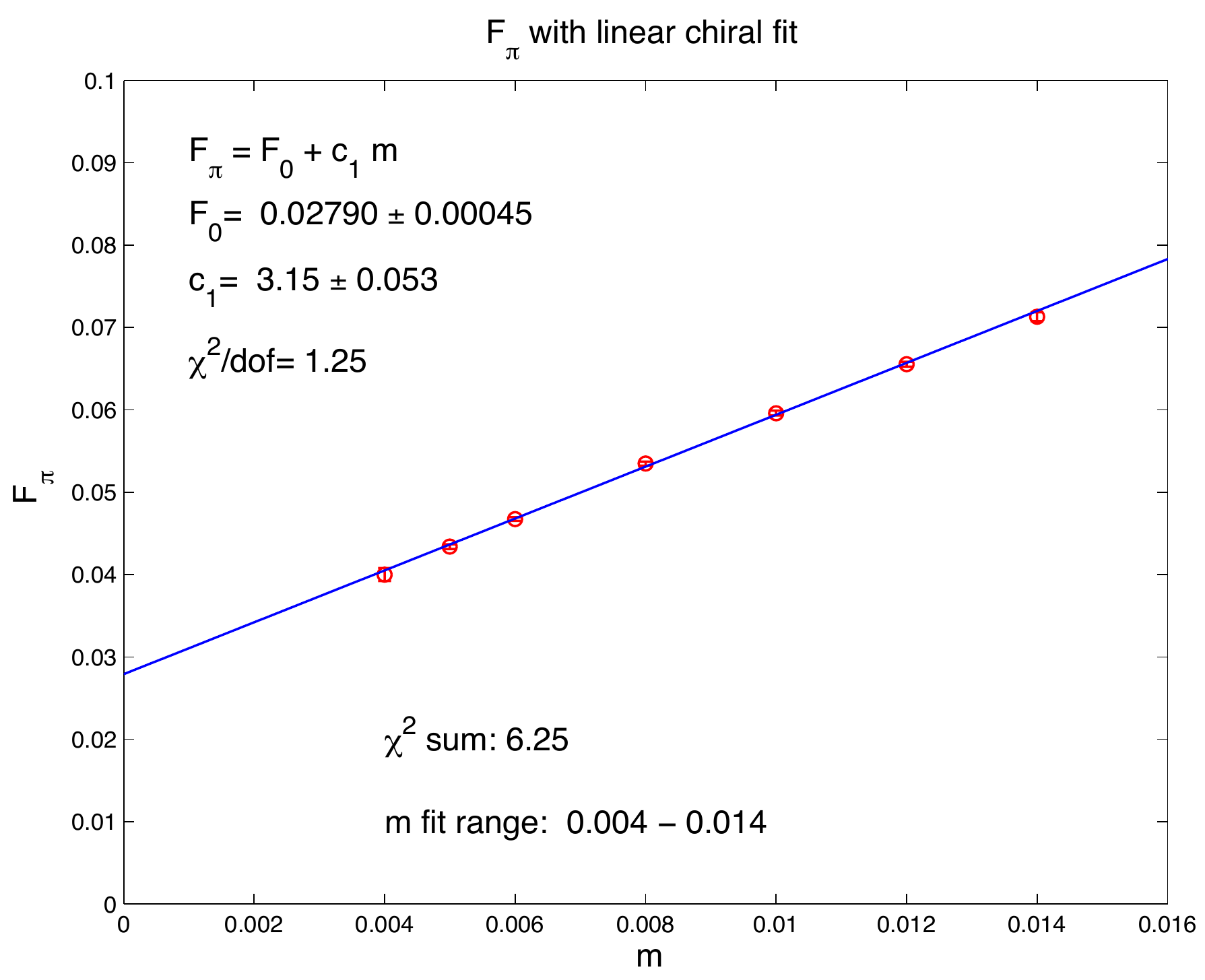}\\
\includegraphics[height=4cm]{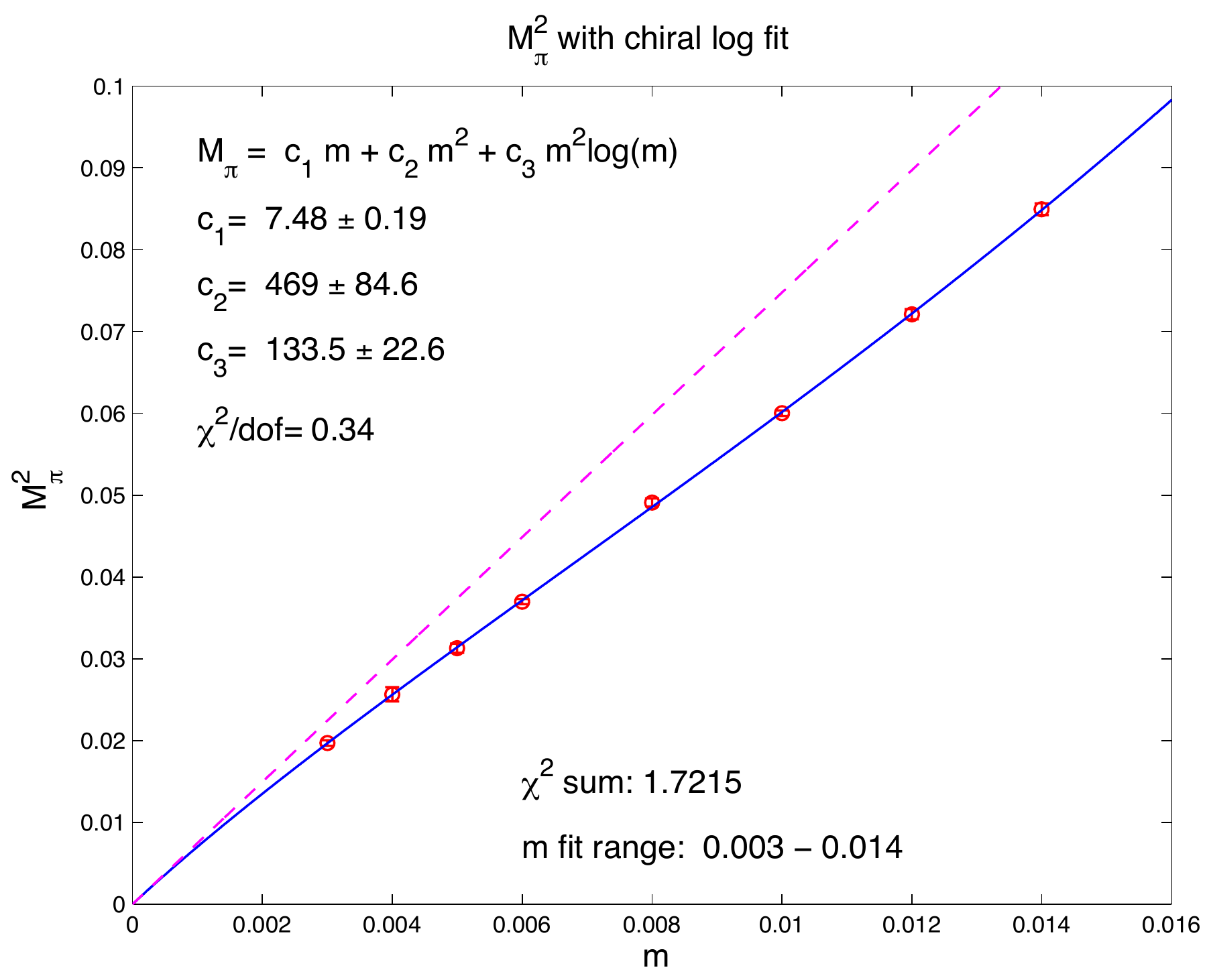}&
\includegraphics[height=4cm]{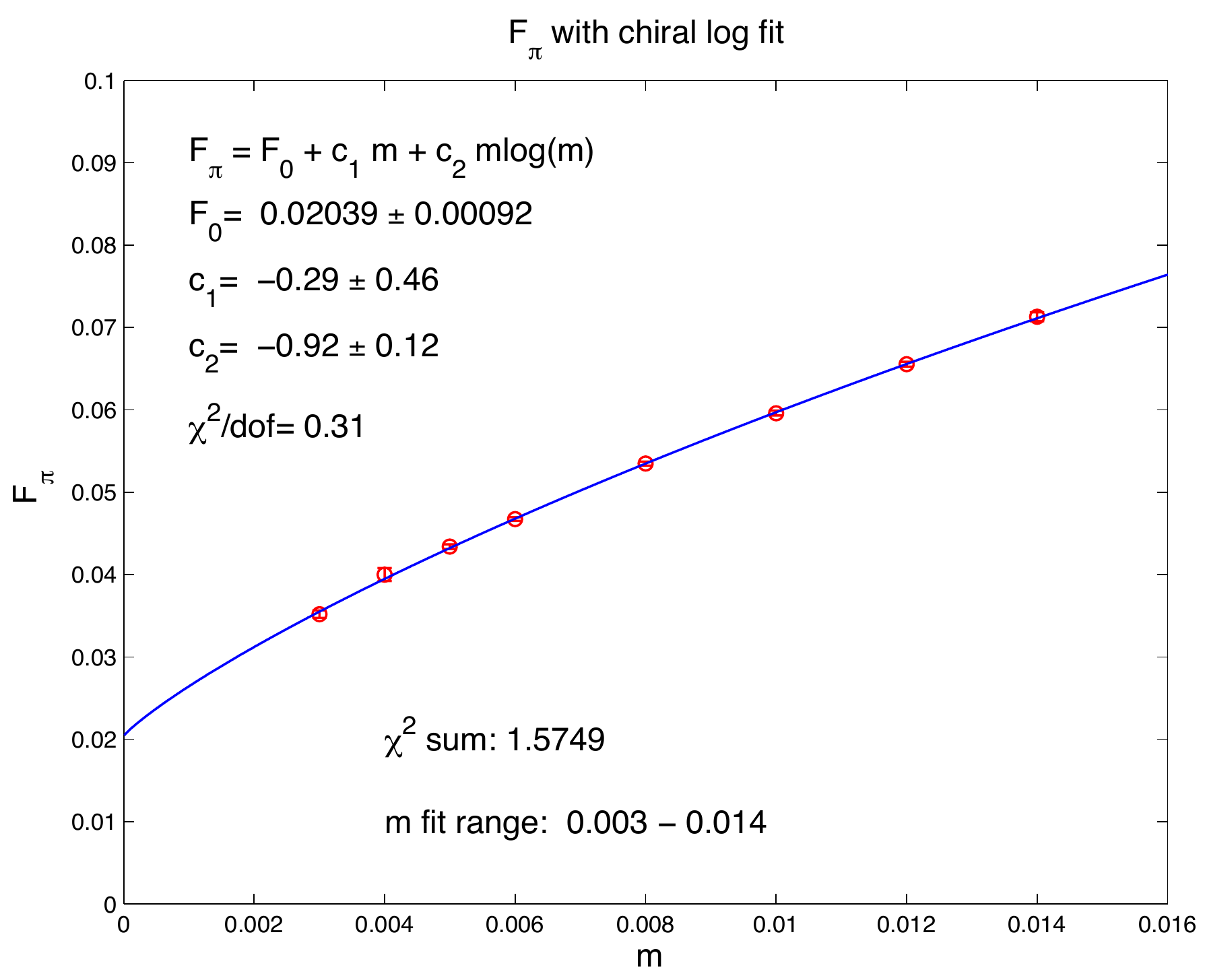}
\end{tabular}
\caption{{\footnotesize Analytic and logarithmic chiral fits for the Goldstone pion and $F_\pi$  are shown for $N_f=2$
sextet simulations with lattice size   $32^3\times 64$ at $\beta=3.2$.
The dashed lines show the linear part of the fits which estimate the $B$-parameter of the chiral Lagrangian. 
The upper two plots are analytic, the lower two plots include the leading
chiral logs. The important role of the $m=0.003$ run is emphasized in the text.}}
\end{center}
\label{fig:sextet-Fpi-Pion}
\vskip -0.2in
\end{figure}
Figure 9 shows the Goldstone pion and $F_\pi$ as a function of the fermion mass $m$ from
 $32^3\times 64$ lattices at the sextet gauge coupling $\beta=3.2$ with some finite volume control from $24^3\times 48$  runs.
 The upper two plots show analytic fits. At $m=0.004$ and higher the finite volume effects are small or negligible. 
 Including the  $m=0.003$ point in the
 analytic fit leads to a substantial increase in $\chi^2$. 
 It remains unclear 
 whether the $m=0.003$ run at $32^3\times 64$ lattice size is compatible with the infinite volume limit or favors an extrapolation.
 Including this point in the logarithmic chiral fit leads to an excellent result. 
 It remains important to resolve the finite
 volume systematics at this fermion mass.
Although we could fit $M_\pi$ and $F_\pi$ with the continuum chiral
logarithm included, the separate sets of $F$ and $B$ from the fits are not quite self-consistent. A combined staggered SU(2)
chiral perturbation theory fit is successful for simultaneous fits of $M_\pi$ and $F_\pi$ with the same pair of $F$ and $B$ values.
The explicit cutoff dependent 
corrections to the $F$ and $B$ parameters would require further testing at weaker gauge coupling.

The sextet non-Goldstone pion spectrum is shown in Figure 10 using the same notation as earlier in the $N_f=12$ model.  
The three states we designate as i5Pion,
ijPion, and scPion do show considerable taste breaking with residual mass in the $m\rightarrow 0$ chiral limit. 
The scPion remains degenerate with the i5Pion and they are  split from the Goldstone pion. 
The ijPion state is further split from the i5Pion as expected. One plot at $\beta=3.25$ also shows that the taste breaking decreases 
at weaker gauge coupling.
The non-Goldstone spectrum is  more QCD-like than in the $N_f=12$ model.

\begin{figure}[htpb]
\begin{center}
\begin{tabular}{cc}
\includegraphics[height=4cm]{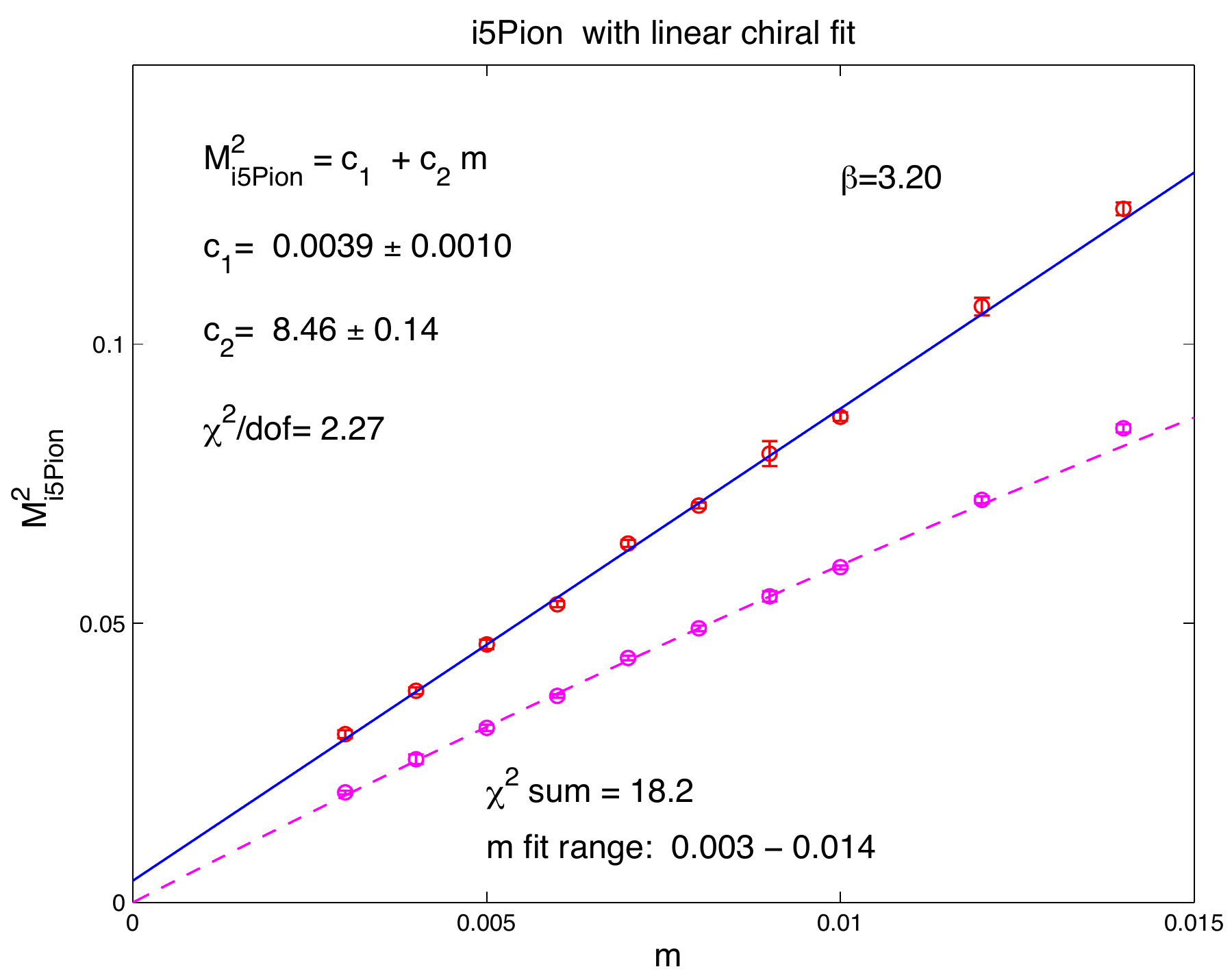}&
\includegraphics[height=4cm]{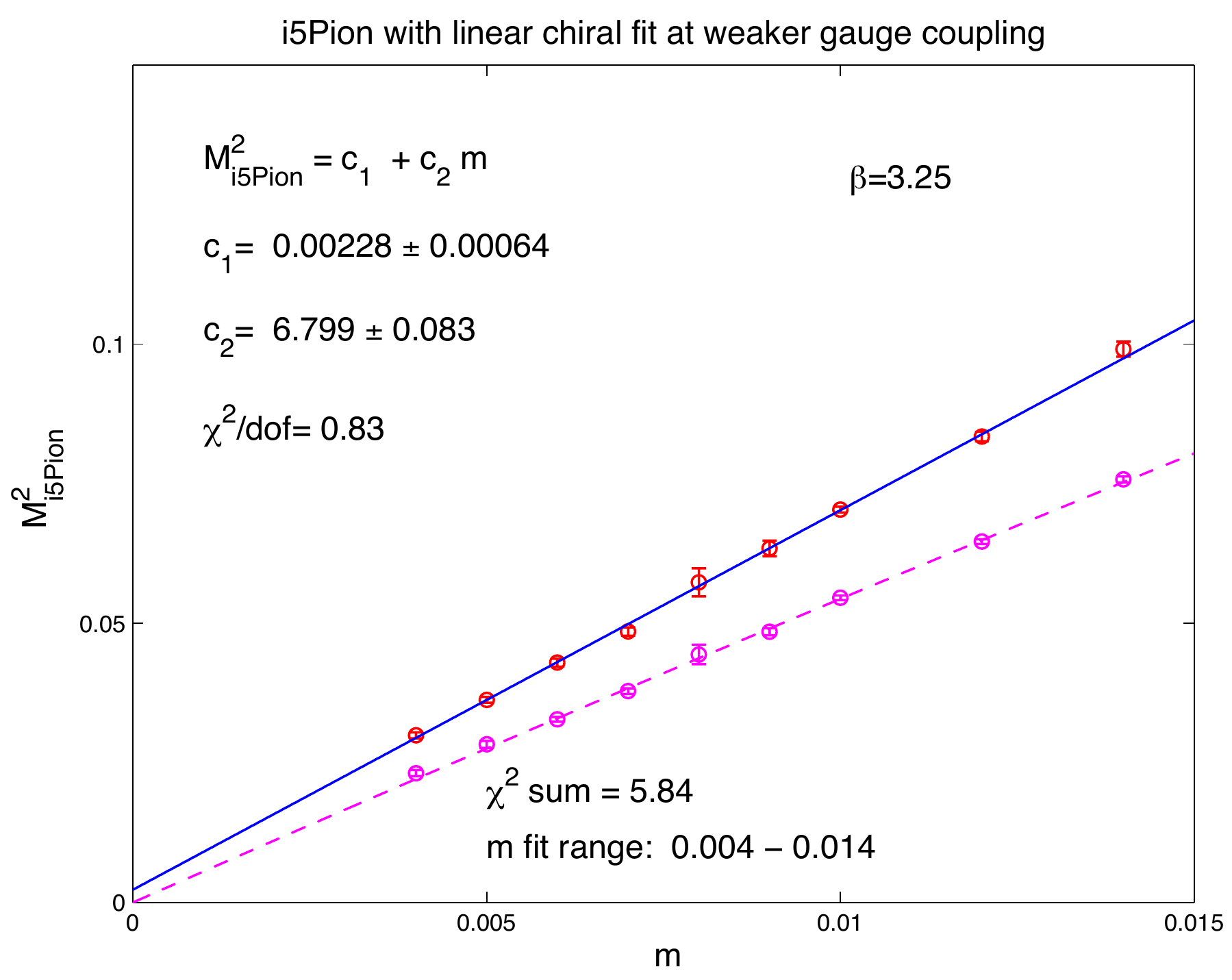}\\
\includegraphics[height=4cm]{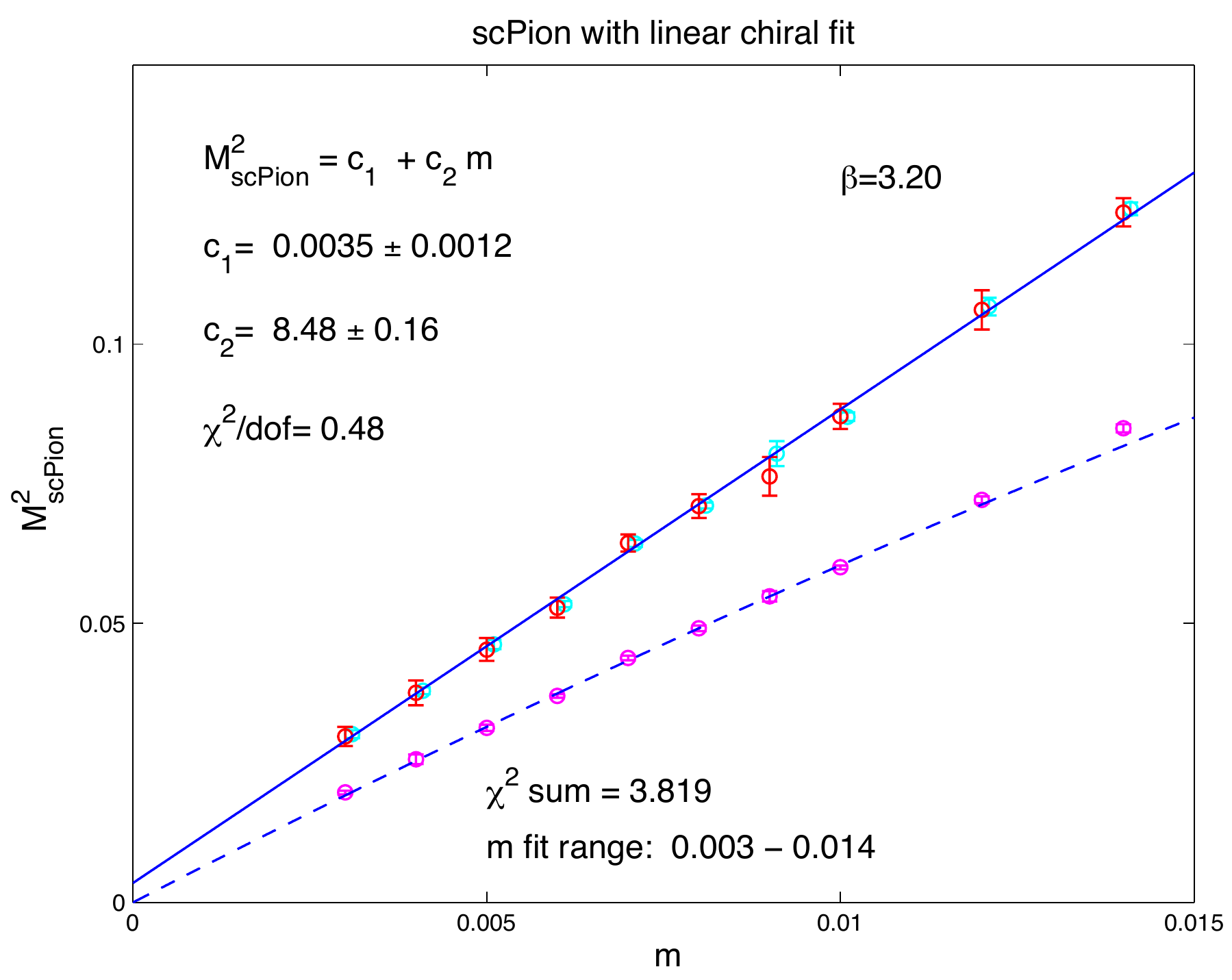}&
\includegraphics[height=4cm]{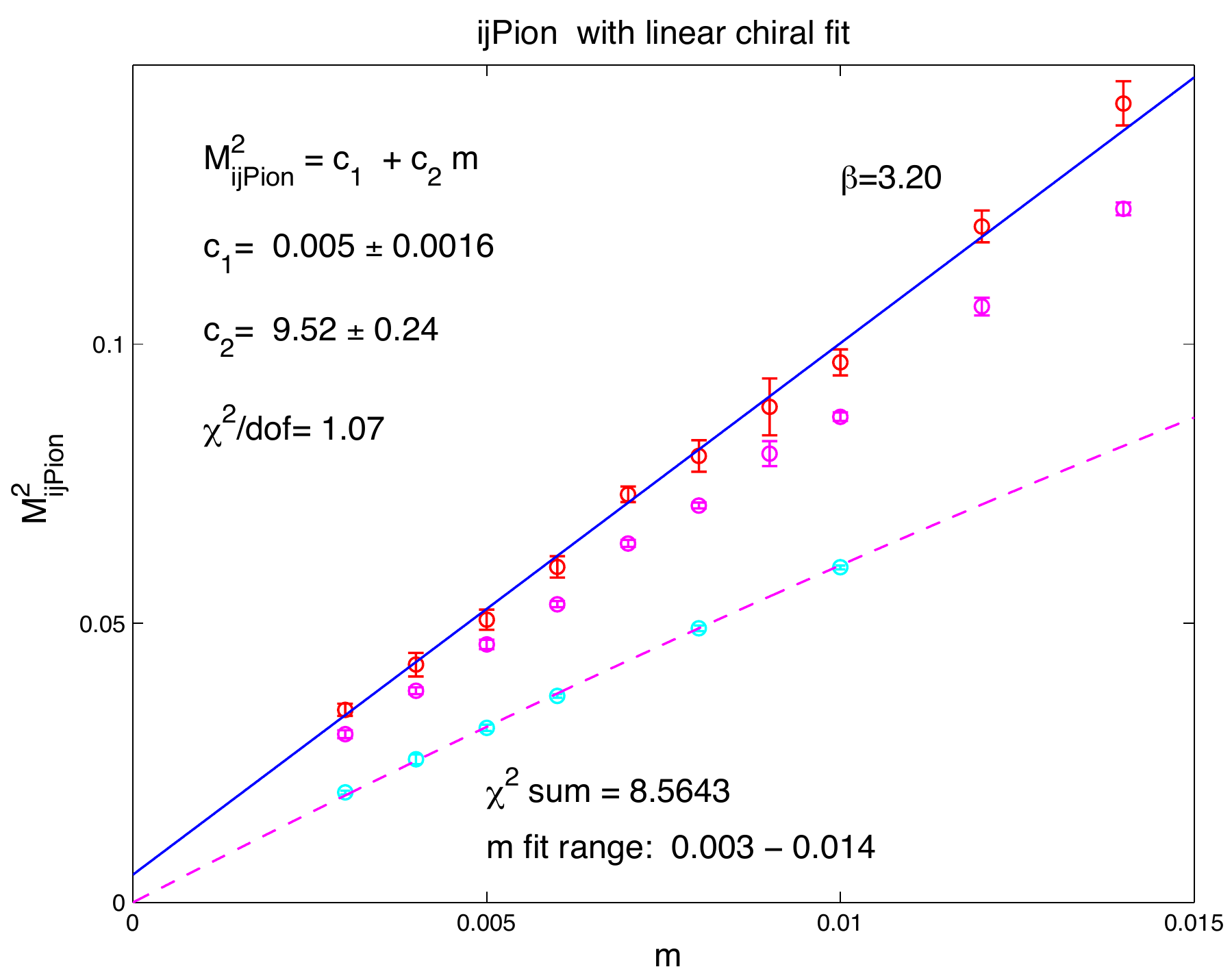}
\end{tabular}
\caption{{\footnotesize Sextet non-Goldstone spectrum is shown for three pion states. The upper left plot at $\beta=3.2$
shows the i5Pion with linear fit and non-zero intercept in the chiral limit. 
%The cyan points are lower mass $24^3\times 48$ runs
%which were not used in the fit. 
The magenta points show the  Goldstone pion and its fit with considerable split from the i5Pion.
The upper right plot is the same plot at $\beta=3.25$ .
%showing stronger volume dependence in the cyan points of the $24^3\times 48$
%runs at lower $m$ values. 
The taste breaking is smaller at weaker gauge coupling. In the bottom left plot the red points 
with the linear fit shows the scPion. The cyan points show the i5Pion which remains degenerate with the scPion. 
The magenta points show
the  Goldstone pion. At the bottom right the red points and the linear fit show the ijPion with magenta 
points showing the i5Pion split downward. The cyan points with its fit show the  Goldstone pion.}}
\end{center}
\label{fig:sextet-goldstone}
\vskip -0.1in
\end{figure}

\subsection{Sextet chiral condensate}

The chiral condensate $\langle \bar{\psi}\psi\rangle$ summed over two flavors
is dominated again by the linear term in $m$ from UV contributions.  The quadratic (or linear) fit
gives a small non-vanishing condensate in the chiral limit which is in the expected
range from the GMOR relation  $\langle\bar{\psi}\psi\rangle=2F^2B$ with the fitted low $F$ value and $B$ estimated from the pion fit
of figure 9. 
For an independent determination, we also studied the chiral condensate operator with the subtracted derivative terms as discussed earlier
in the $N_f=12$ model. The fit to the condensate is shown in Figure 11 with  non-vanishing  chiral limit. 

\begin{figure}[htpb]
\begin{center}
\begin{tabular}{cc}
\includegraphics[height=4cm]{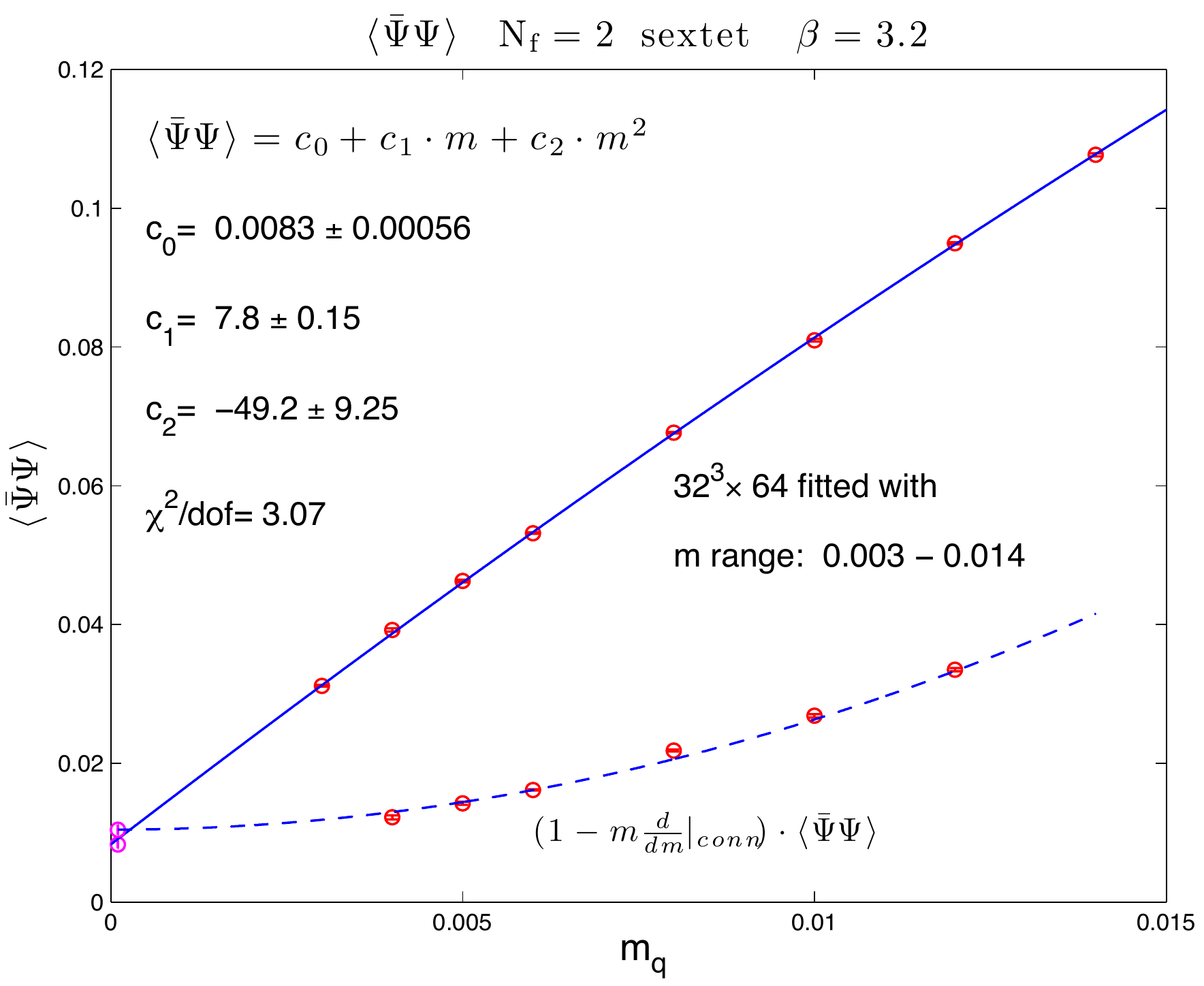}&
\includegraphics[height=4cm]{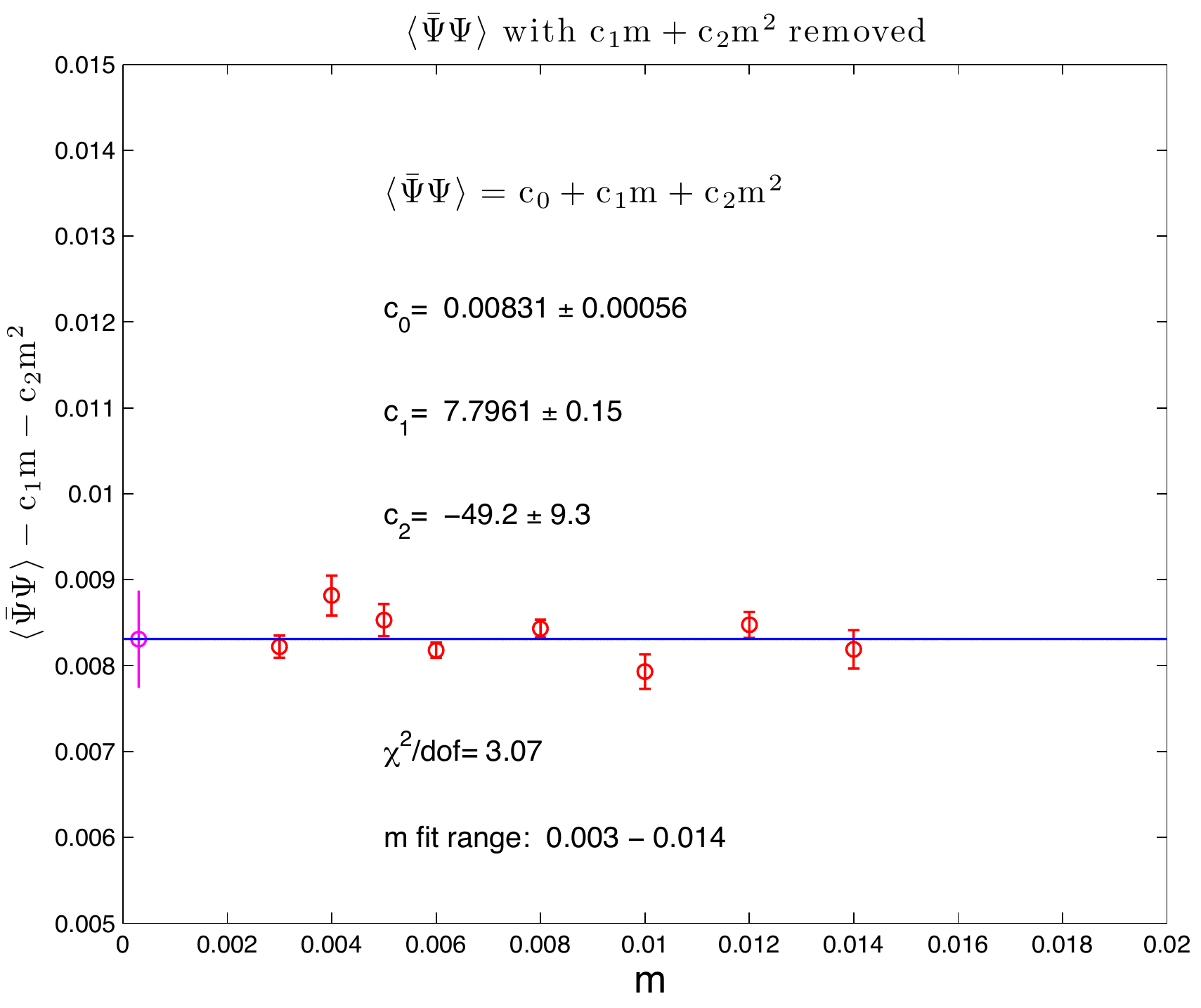}
\end{tabular}
\caption{{\footnotesize The sextet chiral condensate is shown on the left plot with fit and  data points of the 
subtracted derivative. 
The right side shows the fit to  $\langle \bar{\psi}\psi\rangle$ condensate data  after 
the removal of the fitted $c_1m+c_2m^2$ part with fit error on the chiral limit value of  $c_0$ at $m=0$.}}
\end{center}
\label{fig:sextet-PbP}
\vskip -0.2in
\end{figure}

\subsection{ Testing the alternate hypothesis of conformal chiral symmetry in the sextet model}
\begin{figure}[htpb]
\begin{center}
\begin{tabular}{cc}
\includegraphics[height=4cm]{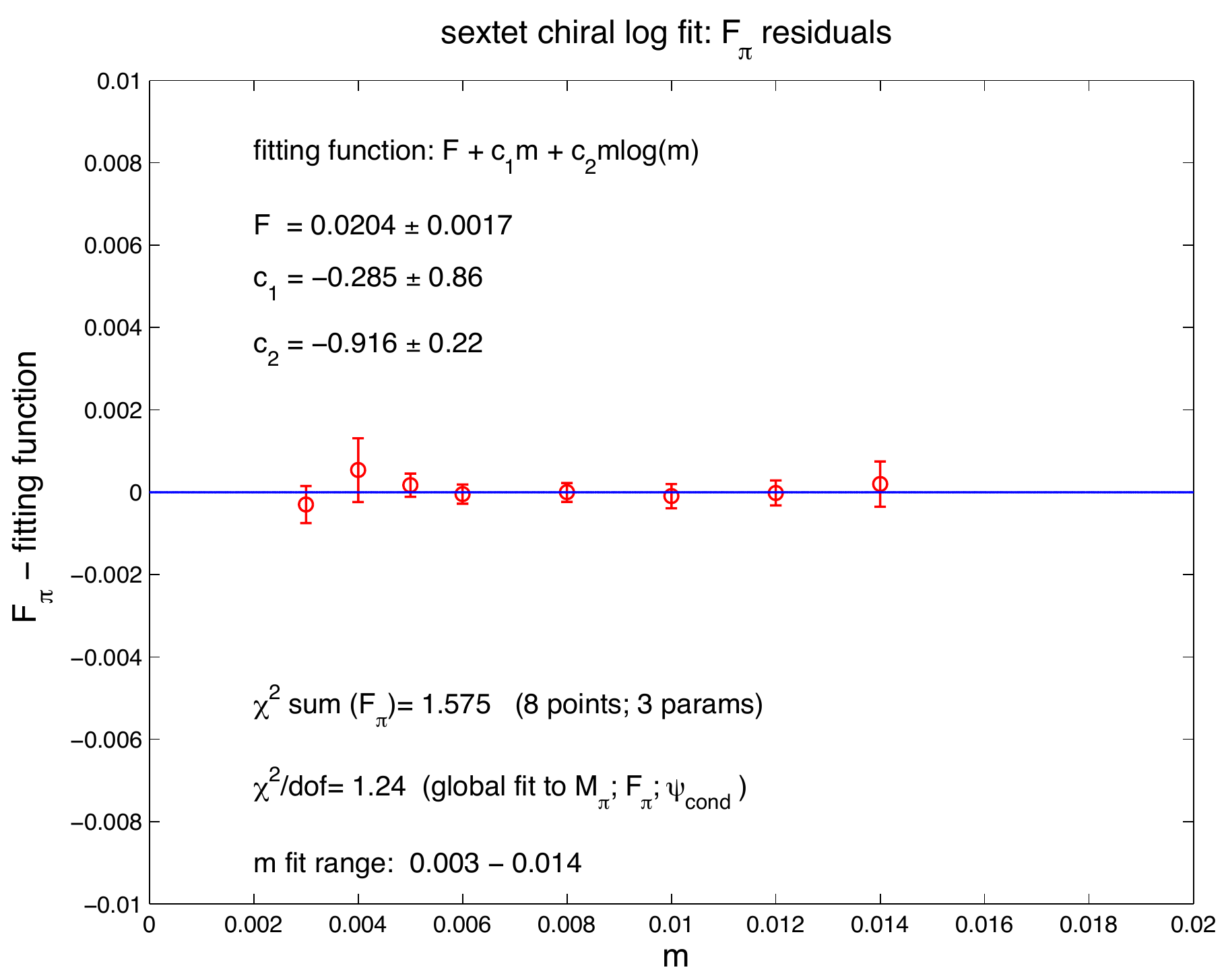}&
\includegraphics[height=4cm]{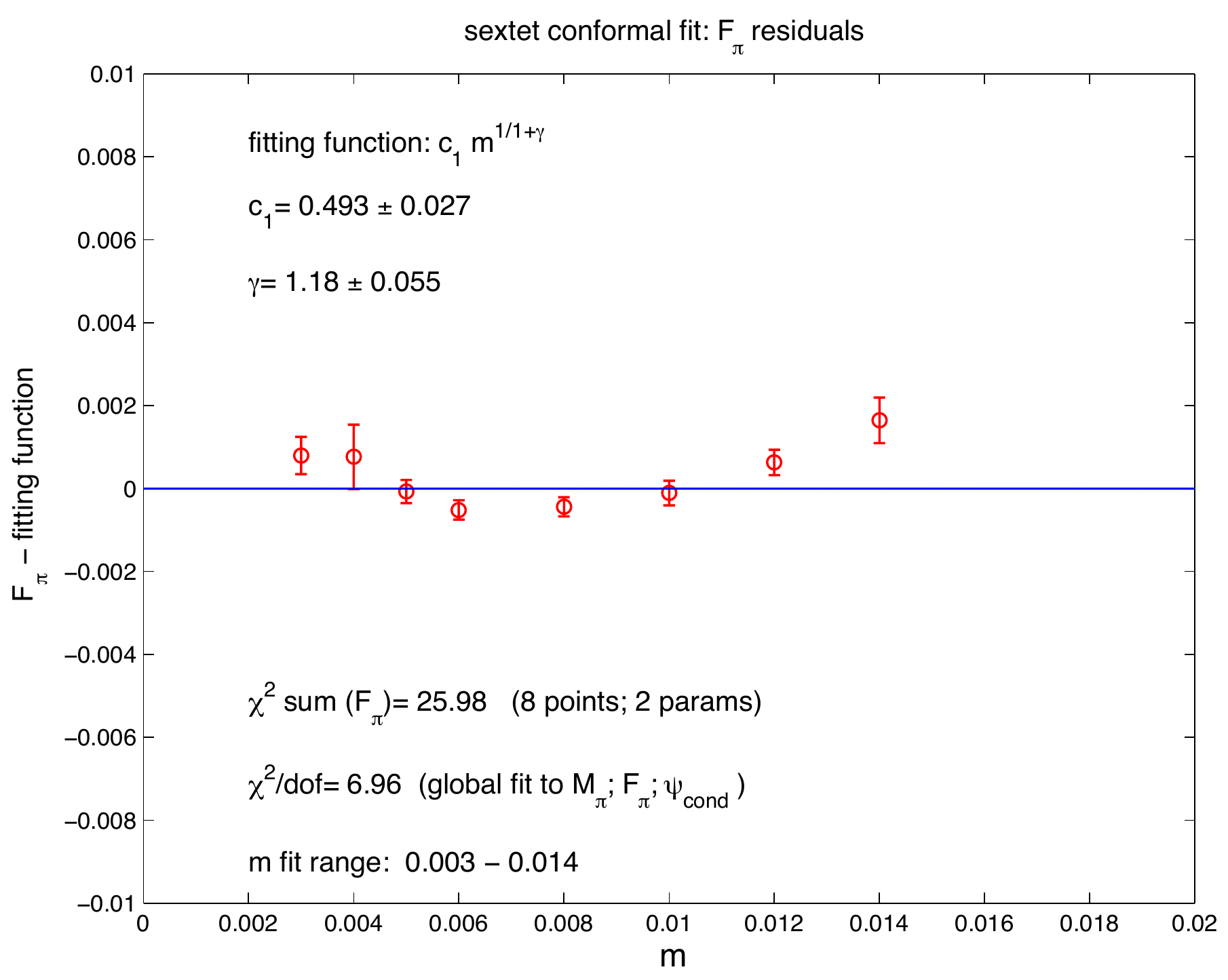}\\
\includegraphics[height=4cm]{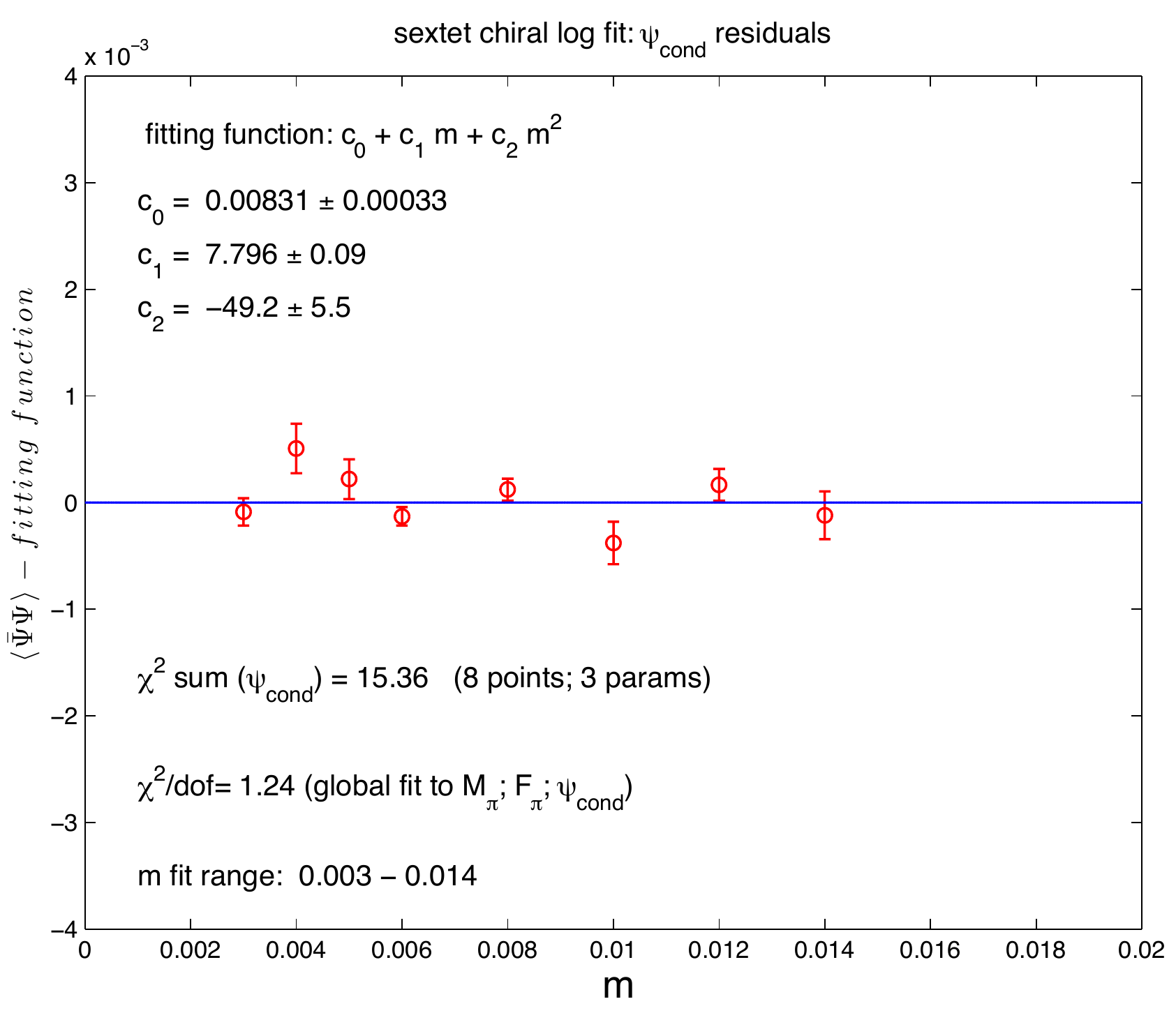}&
\includegraphics[height=4cm]{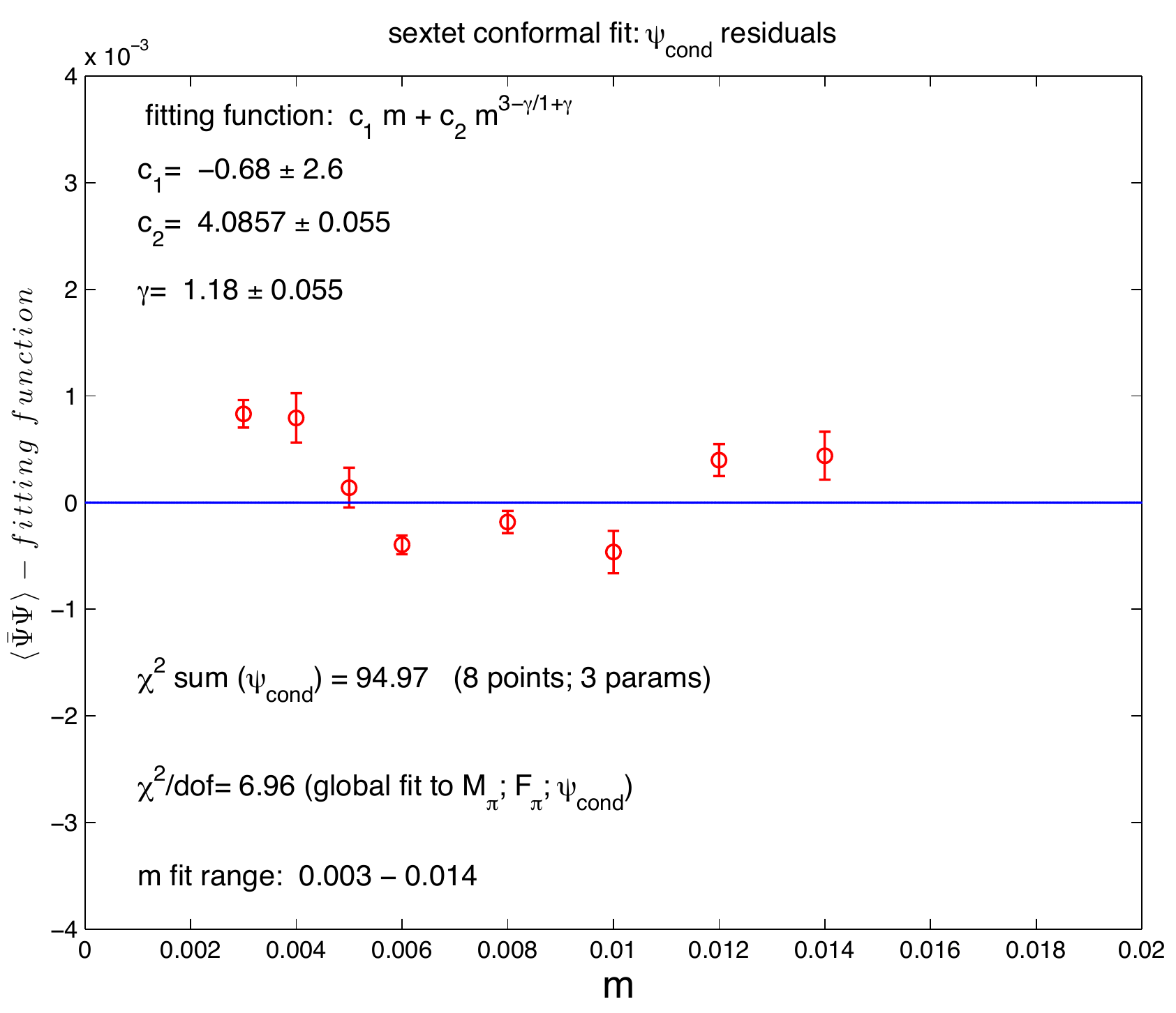}
\end{tabular}
\caption{{\footnotesize The sextet  simultaneous fit in three channels.}}
\end{center}
\label{fig:ConformNf12}
\vskip -0.2in
\end{figure}

All the sextet simulation results we presented favor the chiral symmetry
breaking hypothesis. The pion state is consistent with a vanishing mass in the chiral limit
and easy to fit with a simple quadratic function of the fermion mass. The fundamental 
controlling parameter $F$ of chiral symmetry breaking appears to be significantly non-vanishing in the chiral limit.
The non-vanishing chiral condensate agrees reasonable well with the GMOR relation as 
calculated from the small fitted value of $F$ and $B$ estimated from the pion fit of Figure 9. 
The Higgs ($f_0$) meson, the $\rho$ meson and $A_1$ meson extrapolate 
to non-vanishing masses. 
Applying the
conformal hypothesis to $F_\pi$, $M_\pi$ and the $\langle\bar{\psi}\psi\rangle$ condensate, 
the combined fit gives $\chi^2/{\rm dof}=6.96$ in the $m=0.003-0.014$ fit range
 representing  low level of confidence in the hypothesis.
The combined
chiral fit gives $\chi^2/{\rm dof}=1.24$ in the same range if the chiral log fits
of Figure 9 are used for $M_\pi$ and $F_\pi$. Two comparison fits of this analysis are shown in Figure 12. 
We pointed out earlier that finite volume corrections in the
 $m=0.003$ run cannot be ruled out without further analysis. If we use the $m=0.004-0.014$ range in the
 conformal fits, we get a reduced  $\chi^2/{\rm dof}=3.9$ value and the chiral fit is  $\chi^2/{\rm dof}=1.29$.
 Further work is needed on the $m=0.003$ data.
 
 In summary, our analysis of the sextet model  favors the  $\chi{\rm SB}$ hypothesis with considerable level
 of confidence and disagrees with earlier
 report from ~\cite{Shamir:2008pb} using Wilson fermions in their sextet analysis.
The results disfavoring the conformal hypothesis are not definitive but quite indicative. 
To sharpen the results even more continued work is needed for better control on the systematics at low fermion masses.

%\pagebreak
\acknowledgments{The simulations were performed using computational resources
  at Fermilab and JLab, under the auspices of USQCD and SciDAC, from the
  Teragrid structure and at Wuppertal. We are grateful to Kalman Szabo and Sandor Katz
for their code development. This research is supported by
  the NSF under grants 0704171 and 0970137, by the DOE under grants
  DOE-FG03-97ER40546, DOE-FG-02-97ER25308, by the DFG under grant FO
  502/1 and by SFB-TR/55, and the EU Framework Programme 7 grant
  (FP7/2007-2013)/ERC No 208740.
D.~N. would like to thank the Aspen Center for Physics for invitation to the 2010 BSM summer program.}


\begin{thebibliography}{99}

\footnotesize


%\cite{Appelquist:2007hu}
\bibitem{Appelquist:2007hu}
T.~Appelquist, G.~T.~Fleming and E.~T.~Neil,
%``Lattice Study of the Conformal Window in QCD-like Theories,''
Phys.\ Rev.\ Lett.\  {\bf 100}, 171607 (2008)
%[arXiv:0712.0609 [hep-ph]].
%%CITATION = PRLTA,100,171607;%%

%Us

%\cite{Fodor:2009wk}
\bibitem{Fodor:2009wk}
  Z.~Fodor, K.~Holland, J.~Kuti, D.~Nogradi and C.~Schroeder,
  %``Nearly conformal gauge theories in finite volume,''
  Phys.\ Lett.\ B {\bf 681}, 353 (2009)
  %arXiv:0907.4562
  %%CITATION = ARXIV:0907.4562;%%
  
 %%%%%%%%%%%%%%%%%%%%
% Pallante et al

%\cite{Deuzeman:2009mh}
\bibitem{Deuzeman:2009mh}
  A.~Deuzeman, M.~P.~Lombardo and E.~Pallante,
  %``Evidence for a conformal phase in SU(N) gauge theories,''
  Phys.\ Rev.\  D {\bf 82}, 074503 (2010)
  %[arXiv:0904.4662 [hep-ph]].
  %%CITATION = PHRVA,D82,074503;%%
  
  %%%%%%%%%%%%%%%%%%%%%%%%%%%%%%
% Jin & Mawhinney


%\cite{Jin:2009mc}
\bibitem{Jin:2009mc}
  X.~-Y.~Jin, R.~D.~Mawhinney,
  %``Lattice QCD with 8 and 12 degenerate quark flavors,''
  PoS {\bf LAT2009}, 049 (2009).
  %[arXiv:0910.3216 [hep-lat]].

  
  %%%%%%%%%%%%%%%%%%%%%%%%%%%%%%%%
% Anna Hasenfratz et al

%\cite{Hasenfratz:2010fi}
\bibitem{Hasenfratz:2010fi}
  A.~Hasenfratz,
  %``Conformal or Walking? Monte Carlo renormalization group studies of SU(3) gauge models with fundamental fermions,''
  Phys.\ Rev.\  {\bf D82}, 014506 (2010).
  %[arXiv:1004.1004 [hep-lat]].
%%%%%%%%%%%%%%%%%%%%%%

%\cite{Morningstar:2003gk}
\bibitem{Morningstar:2003gk}
  C.~Morningstar and M.~J.~Peardon,
  %``Analytic smearing of SU(3) link variables in lattice QCD,''
  Phys.\ Rev.\  D {\bf 69}, 054501 (2004)
  %[arXiv:hep-lat/0311018].
  %%CITATION = PHRVA,D69,054501;%%
%%%%%%%%%%%%%%%%%%%%%%%%%%%%%%%%%%  


%\cite{Aoki:2005vt}
\bibitem{Aoki:2005vt}
  Y.~Aoki, Z.~Fodor, S.~D.~Katz, K.~K.~Szabo,
  %``The Equation of state in lattice QCD: With physical quark masses towards the continuum limit,''
  JHEP {\bf 0601}, 089 (2006).
%  [hep-lat/0510084].
%%%%%%%%%%%%%%%%%%%%%%%%%%%%%%%%%%% 


\bibitem{Durr:2010aw}
  S.~Durr, Z.~Fodor, C.~Hoelbling, S.~D.~Katz, S.~Krieg, T.~Kurth, L.~Lellouch, T.~Lippert {\it et al.},
  %``Lattice QCD at the physical point: Simulation and analysis details,''
  [arXiv:1011.2711 [hep-lat]].

%%%%%%%%%%%%%%%
% Catterall et al

%\cite{Catterall:2007yx}
\bibitem{Catterall:2007yx}
  S.~Catterall and F.~Sannino,
  %``Minimal walking on the lattice,''
  Phys.\ Rev.\  D {\bf 76}, 034504 (2007)
  %[arXiv:0705.1664].
  %%CITATION = PHRVA,D76,034504;%%


%%%%%%%%%%%%%%%%%

% Rummukainen et al

%\cite{Hietanen:2009az}
\bibitem{Hietanen:2009az}
  A.~J.~Hietanen, K.~Rummukainen and K.~Tuominen,
  %``Evolution of the coupling constant in SU(2) lattice gauge theory with two
  %adjoint fermions,''
  Phys.\ Rev.\  D {\bf 80}, 094504 (2009)
  %[arXiv:0904.0864 [hep-lat]].
  %%CITATION = PHRVA,D80,094504;%%

%%%%%%%%%%%%%%%%%%%%

% Del Debbio et al

\bibitem{DelDebbio:2010hx}
  L.~Del Debbio, B.~Lucini, A.~Patella, C.~Pica, A.~Rago,
  %``The infrared dynamics of Minimal Walking Technicolor,''
  Phys.\ Rev.\  {\bf D82}, 014510 (2010).
%  [arXiv:1004.3206 [hep-lat]].

%\bibitem{DeGrand:2009hu}
%  T.~DeGrand,
  %``Finite-size scaling tests for SU(3) lattice gauge theory with color sextet fermions,''
%  Phys.\ Rev.\  {\bf D80}, 114507 (2009).
%  [arXiv:0910.3072 [hep-lat]].
%%%%%%%%%%%%%%%%%%%%%%%%%%%%%%%
% DeGrand et al

%\cite{Shamir:2008pb}
\bibitem{Shamir:2008pb}
  Y.~Shamir, B.~Svetitsky and T.~DeGrand,
  %``Zero of the discrete beta function in SU(3) lattice gauge theory with color
  %sextet fermions,''
  Phys.\ Rev.\  D {\bf 78}, 031502 (2008)
  %[arXiv:0803.1707 [hep-lat]].
  %%CITATION = PHRVA,D78,031502;%%


%%%%%%%%%%%%%%%%%%%%%%%%%%%%%%%%%%%
% Sinclair & Kogut

%\cite{Kogut:2010cz}
\bibitem{Kogut:2010cz}
  J.~B.~Kogut, D.~K.~Sinclair,
  %``Thermodynamics of lattice QCD with 2 flavours of colour-sextet quarks: A model of walking/conformal Technicolor,''
  Phys.\ Rev.\  {\bf D81}, 114507 (2010).
  %[arXiv:1002.2988 [hep-lat]].


%%%%%%%%%%%%%%%%%%%%%%%%%%%%%%%%%%%
% Itou et al

%\cite{Bilgici:2009kh}
\bibitem{Bilgici:2009kh}
  E.~Bilgici {\it et al.},
  %``A new scheme for the running coupling constant in gauge theories using
  %Wilson loops,''
  Phys.\ Rev.\  D {\bf 80}, 034507 (2009)
  %[arXiv:0902.3768 [hep-lat]].
  %%CITATION = PHRVA,D80,034507;%%

%%%%%%%%%%%%%%%%%%%%%%%%%%%%%%%%%%%%%%
% Yamada et al

%\cite{Yamada:2009nt}
\bibitem{Yamada:2009nt}
  N.~Yamada, M.~Hayakawa, K.~-I.~Ishikawa {\it et al.},
  %``Study of the running coupling constant in 10-flavor QCD with the Schrodinger functional method,''
  PoS {\bf LAT2009}, 066 (2009).
  %[arXiv:0910.4218 [hep-lat]].



\end{thebibliography}
\end{document}